\documentclass[12pt]{article}
\usepackage{latexsym}
\newcommand{\be}{\begin{equation}}
\newcommand{\ee}{\end{equation}}
\newcommand{\bq}{\begin{eqnarray}}
\newcommand{\eq}{\end{eqnarray}}
\newcommand{\ids}{\int d^{2}\!\sigma}
\newcommand{\pdp}{\partial_{+}}
\newcommand{\dm}{\partial_{-}}
\newcommand{\ns}{\nu\sigma}
\newcommand{\dsm}{\partial_{\sigma_{-}}}
\newcommand{\ssp}{\sigma_{+}}
\newcommand{\sm}{\sigma_{-}}
\newcommand{\vsm}{v_{-}(\sigma_{-})}

\newcommand{\dms}{\frac{\delta}{\delta X^{\mu\sigma}}}

\newcommand{\idsp}{\int d^{2}\!\sigma'}
\newcommand{\lt}{\lambda\tau}
\newcommand{\vessup}{\frac{v(\sigma_{+})-v(\sigma'_{+})}{\sigma_{+}
-\sigma'_{+}}}

\begin{document} 
\begin{titlepage}
\today          \hfill 
\begin{center}
\hfill    LBL-40936 \\
          \hfill    UCB-PTH-97/53 \\

\vskip .5in

{\large \bf String Field Equations From \\
Generalized Sigma Model II}
\footnote{This work was supported in part by the Director, Office of 
Energy Research, Office of High Energy and Nuclear Physics, Division of 
High Energy Physics of the U.S. Department of Energy under Contract 
DE-AC03-76SF00098 and in part by the National Science Foundation under 
grant PHY-95-14797.}

\vskip .5in
Korkut Bardakci\footnote{e-mail: kbardakci@lbl.gov }
\vskip .5in

{\em Theoretical Physics Group\\
    Lawrence Berkeley Laboratory\\
      University of California\\
    Berkeley, California 94720}
\end{center}

\vskip .5in

\begin{abstract}

We improve and extend a method introduced in an earlier paper for deriving
string field equations. The idea is to impose conformal invariance on
a generalized sigma model, using a background field method that ensures
covariance under very general non-local coordinate transformations. The method
is used to derive the free string equations, as well as the interacting
 equations for the graviton-dilaton system. The full interacting string field
 equations  derived by this method should be manifestly
background independent. 

\end{abstract}
\end{titlepage}

\newpage
\renewcommand{\thepage}{\arabic{page}}
\setcounter{page}{1}
\noindent{\bf 1. Introduction}
\vskip 9pt

This paper is the follow up of an earlier
 paper with the same title [1].
 The basic idea of both papers is to derive the dynamical equations
satisfied by the string states by requiring a sigma model  on the
world sheet to be conformally invariant. This idea has a long history,
going back to some early papers [2-6],
 where the equations satisfied by the massless particles in the
spectrum of the string were derived by demanding conformal invariance of
the effective action in the one loop approximation.
 These early efforts dealt with only
renormalizable sigma models, which restricted the scope of their investigation
to the dynamics of the massless states. In order to incorporate the dynamics
of the massive levels of the string, one has to start with the most general
non-renormalizable sigma model, subject only to some general requirements
of invariance. A  method of imposing conformal invariance on a
a general sigma model was proposed by Banks and Martinec [7], who
introduced an explicit cutoff and used the Wilson type renormalization
group equations [8,9]. This approach was further developed
and was used to derive the tree level closed bosonic string amplitudes
by Hughes, Liu and Polchinski[10] and others [11-14].
 The idea behind this
method is to cancel the conformal anomalies due to the quantum corrections
against the classical violation of the conformal symmetry due to
 the presence of nonrenormalizable terms in the action. Despite its success
in reproducing string amplitudes, this approach  suffers from some
drawbacks, among them lack of a sufficiently powerful gauge invariance
to eliminate all the spurious states [10]. Another disadvantage
of this approach is the absence of manifest covariance under 
redefinitions of the target space coordinate $X(\sigma)$. In fact, it will
become clear later on that these problems are related; the spurious states
are absent in a manifestly covariant treatment.

In the reference cited above [1],
 we proposed a new method for deriving the
string field equations, by combining the advantageous features of
both the earlier work on the sigma model [2-6],
 and  of the Wilson renormalization group approach [7,10].
 The starting point was
  the most general nonrenormalizable sigma model
on the world sheet, subject only to two dimensional Poincare invariance.
The basic idea was again to cancel the quantum conformal anomaly by the
terms in the action that violate conformal invariance classically. This was
done by first computing the one loop effective action with an explicit
 cutoff, and then by requiring the effective action to be invariant
under conformal transformations. The main goal of the paper was to carry
out the calculation of the effective action using a background field
 method [3,15],  which is covariant under field
 transformations. The  transformations in question included not only
the local diffeomorphisms of general relativity, but also non-local ones
with derivatives with respect to the world sheet coordinates (see eq.(4)).
This posed the problem of finding a suitable covariant metric in order to
be able to use the tools of differential geometry. Such a metric can easily be
extracted from the action when only covariance  under the local transformations
is required; however, when non-local transformations are also included,
the problem becomes  difficult. In reference [1],
partial progress was made in this direction by using an expansion in the
slope parameter; however, only the equations for the first few levels
of the string could be derived by this method. Even then,
 the left-right nonsymmetric string could not be treated .

In the present paper, we show how to overcome all of the difficulties 
encountered in the earlier work. The main new idea is to forget about the
metric and introduce the connection as a completely independent field.
It is also necessary to introduce a vector field which generates the
conformal transformations (conformal Killing vector), again as an
independent quantity. This means that we make no a priori commitment about
the metric and the conformal Killing vector, but instead, we let the
the equations resulting from conformal invariance (the RG equations)
decide the issue. However, one encounters several problems in applying
these equations: They are explicitly cutoff dependent and also they do not seem
powerful enough to fix the connection and the Killing vector completely.
The first problem is not really serious;
 it turns out
that almost all of the cutoff dependendence factorizes, leaving behind
cutoff independent equations. The only exception is a set of terms with
logarithmic dependence on the cutoff, and these can be eliminated by slope
renormalization. This is then the only renormalization needed to render
the theory finite. As for the second problem, it is
true that the connection and the Killing vector remain mostly undetermined;
 however, this does not mean that the RG equations contain
no useful information. A subset of the equations turn out to be independent
of the connection and the Killing vector, and these equations are then the
candidates for the string field equations. A major part of this paper is
devoted to working out the consequences of this idea to see whether it
actually leads to the correct string equations. This comparison is done
in two different cases: First, the linearized form of the RG equations are
shown to be equivalent to the the well known free string equations. Also,
going beyond the linear approximation, the interacting graviton-dilaton
equations  come out correctly. 

The paper is organized as follows: In section 2, we review the version of the
 RG equations derived in [1], and we rewrite them in a form
convenient for future applications. We also discuss in some detail the
heat kernel method, the regularization scheme we use in this paper. It has
a number of advantages over the explicit cutoff used in [1].
In  sections 3 and 4, the linearized RG equations are applied to the 
massless and the first massive levels of the string. There are several
reasons for
considering these special cases before embarking on the general problem.
The same special cases were considered in [1]; here we show
how the present treatment overcomes the difficulties encountered there.
Also, many of the important features of the general problem are already
present in these special cases, and working them out in detail should be
helpful. For example, one can easily verify that the cutoff dependence
of the equations causes no problems. Also, it is instructive to see
that the covariant treatment helps to eliminate several spurious states
and the resulting spectrum is then in agreement with the string spectrum.
 In section 5, we apply the linearized RG equations to an
arbitrary string state, and we show that they can be written in a
compact form as a single equation
  using the standard operator formalism familiar from
string theory. Section 6 is devoted to establishing the equivalence of this
equation to the standard equations satisfied by the free string. Finally,
we go beyond the linear approximation in section 7
 by applying the full non-linear RG equations to the dilaton-graviton
system, and we show that the resulting equations are the correct ones.

By working out these examples, we hope to have shown that
 the approach to string
field equations proposed here is both correct and useful.
 As a future project, it seems
quite feasible to derive the full set of interacting equations in the
operator formalism of section 5. The main motivation for doing this is
the realization that these equations should be manifestly
 background independent.
Although initially the calculations are done in the framework of
 an expansion around the flat background, using the methods of sections
5 and 7, one should be able to sum the series and get rid of the background
dependence. Lack of manifest background independence is a problem shared
by many different approaches to string field theory, including the BRST
formalism [16-19].
 In addition to background independence, the field equations derived by
the present method will also be
invariant under non-local field transformations mentioned earlier. It has been
suspected for a long time that string theory has a large class of as yet
undiscovered hidden symmetries, and that these symmetries may be 
important in understanding string dynamics. For example, duality
symmetries [20], which have attracted much attention recently, may be
the manifestations of a much bigger hidden symmetry. In any case, 
 any new approach to string theory will hopefully deepen our understanding
of it.

\noindent {\bf 2. One Loop RG Equations}
\vskip 9pt

We start this section with a brief review of the
 one loop RG equations derived in [1].
 The starting point is a two dimensional
 action $S$ which describes the world sheet structure of an interacting bosonic
string theory. The only requirement on this action is two dimensional Lorenz
invariance, other than that, it is the most general local non-renormalizable
action constructed from the string coordinate $X^{\mu\sigma}\equiv
 X^{\mu}(\sigma)$. All of the computations of this paper will be carried out
in a flat Minkowski background, accordingly, the action is
 split  into free and interacting parts:
\bq
S & =& S^{(0)}+S^{(1)}, \nonumber \\
S^{(0)}& =& \ids \:\pdp\! X^{\mu\sigma}\:\dm\!X^{\ns} \eta_{\mu\nu},\nonumber \\
S^{(1)}& =& \ids\left(\Phi(X(\sigma))+\tilde{h}_{\mu\nu}(X(\sigma))
\pdp\!X^{\mu\sigma} \dm\!X^{\ns}+\cdots \right),
\eq
where $\eta_{\mu\nu}$ is the flat Minkowski metric, $\Phi$ is the
 tachyon field, and $\tilde{h}_{\mu\nu}$ is related to
 the gravitational metric $g_{\mu\nu}$ and the
 antisymmetric tensor $B_{\mu\nu}$ through
\be
\tilde{g}_{\mu\nu}=\eta_{\mu\nu}+\tilde{h}_{\mu\nu},\;\;\; g_{\mu\nu}=\frac{1}
{2}(\tilde{g}_{\mu\nu}+\tilde{g}_{\nu\mu}),\;\;\;B_{\mu\nu}=\frac{1}{2}(\tilde{
g}_{\mu\nu}-\tilde{g}_{\nu\mu}),
\ee
and $\pdp \equiv \partial_{\sigma_{+}}$ and $\dm \equiv \partial_{\sigma_{-}}$
 are derivatives with respect to the world sheet coordinates
$$
\ssp=\frac{1}{2}(\sigma_{0}+i\sigma_{1}),\;\;\;\sm=\frac{1}{2}(\sigma_{0}-
\sigma_{1}). 
$$
The dots represent higher levels which contain more derivatives with
respect to $\sigma$.
Eq.(1) is a quasi-local expansion of the action
 in the derivatives of the coordinate $X^{\mu}(\sigma)$; the
fields are local functions of $X^{\mu}(\sigma)$, as opposed to functionals.
 Non-locality is introduced gradually through higher powers of
 $\partial_{\pm}X^{\mu\sigma}$.
World sheet Lorentz invariance requires equal numbers of $\partial_{+}$ and
$\partial_{-}$. The presence of higher derivatives makes
 the model unrenormalizable, and a cutoff is needed to define it. Another way
to organize this expansion is to classify the terms according to their
classical conformal dimension, which is the naive dimension associated with
the scaling of  $\sigma$: Each derivative with respect to $\sigma$ adds a
unit to the classical conformal dimension. We note that the action is not
classically conformal invariant.

The invariance properties of the model will play an important role.
 The action of eq.(1) is invariant if a total derivative
 is added to the integrand, setting
$$
S=\ids\: I(\sigma) .
$$
the action is invariant under
\be
I\rightarrow I+ \pdp I_{-}(\sigma)+ \dm I_{+}(\sigma).
\ee
Later, we will see that in the string language, this corresponds to invariance
under adding spurious states generated by the application of the Virasoro
operators $L_{-1}$ and $\bar{L}_{-1}$ to the physical states.
 We will call this a linear gauge transformation. In addition
 to these invariances, which follow automatically from the
 definition of the action, we will impose invariance under
 the infinitesimal coordinate transformations
\be
X^{\mu\sigma}\rightarrow X^{\mu\sigma}+f^{\mu}(X(\sigma))+
f^{\mu}_{\nu\lambda}(X(\sigma)) \pdp X^{\nu\sigma}\dm X^{\lambda\sigma}
+\cdots
\ee
where f's are arbitrary local functions of $X(\sigma)$. The first function
 $f^{\mu}$ corresponds to the local diffeomorphisms of general relativity, so 
it ensures the imbedding of gravity into the model.
We shall see later that the transformations with 
higher derivatives eliminate spurious states.

Finally, we would like the model to be conformally invariant. In the flat 
world sheet formulation we are using, the two sets of infinitesimal conformal
transformations are given by
\be
\ssp \rightarrow \ssp + v_{+}(\ssp), \;\;\; \sm \rightarrow \sm + \vsm.
\ee
The following operators, acting on the coordinates, generate these
 transformations:
\be
\delta_{v_{\pm}}= \ids\: v_{\pm}(\sigma_{\pm})\:
 \partial_{\pm}\!X^{\mu\sigma} \dms.
\ee
However, these generators do not transform properly under the coordinate
transformations given by eq.(4). To ensure proper transformation properties,
$\partial_{\pm}\!X^{\mu\sigma}$ in eq.(6) should be replaced
 by a vector (Killing vector):
\bq
\delta_{v_{\pm}}&=& \ids F^{\mu\sigma}_{v_{\pm}}(X) \dms,\nonumber \\
F^{\mu\sigma}_{v_{\pm}}& =&  v_{\pm}(\sigma_{\pm}) \partial_{\pm}\!
X^{\mu\sigma} +
\idsp v_{\pm}(\sigma'_{\pm}) f^{\mu\sigma}_{\sigma'}(X).
\eq
Here, $f^{\mu\sigma}_{\sigma'}$ is
 introduced so that $F^{\mu\sigma}_{v_{\pm}}$ will
 transform like a contravariant vector in the indices $\mu\sigma$ under the
transformations of eq.(4). This then guarantees that conformal invariance is
coordinate independent. To start with, $f^{\mu\sigma}_{\sigma'}$ will be left
arbitrary, and it will eventually be fixed by the string field equations.

The string field equations can be derived [10]
 by requiring the conformal invariance of the string theory based
on the action S (eq.(1)). This action is not even classically conformal
 invariant
 as it stands; the tachyon field and the fields corresponding to
massive levels violate classical conformal invariance. Quantum mechanically,
there is a further violation (anomaly) coming from higher order graphs.
Conformal invariance can be restored by cancelling the classical terms
against the quantum anomaly; the resulting conditions are then the string
field equations. Below, we write down the version of
these equations derived in [1]:
\be
E_{G}+E_{M}=0,
\ee
where,
\be
E_{G}=\left(F^{\mu\sigma}_{v}\dms+
\delta_{\Lambda}\right)\left(bS-\frac{1}{2}
Trlog (G)\right),
\ee
and
\bq
E_{M}& =& \frac{1}{2}\left(-F^{\lt}_{v}\:\frac{\delta G^
{\mu\sigma,\mu'\sigma'}}
{\delta X^{\lt}}+\frac{\delta F^{\mu\sigma}_{v}}{\delta X^{\lt}}\:G
^{\lt,\mu'\sigma'}
+\frac{\delta F^{\mu'\sigma'}_{v}}{\delta X^{\lt}}\:G^{\mu\sigma,\lt}-
\delta_{\Lambda}
G^{\mu\sigma,\mu'\sigma'}\right) \nonumber \\
&\times & M_{\mu\sigma,\mu'\sigma'}.
\eq

Let us define the expressions that appear in the equation above. 
   $F$ and $S$ were discussed earlier, ``b'' is the slope parameter, and
 $\delta_{\Lambda}$ involves the variation of the cutoff and it will be
explained when we discuss the cutoff procedure. 
 The``supermetric'' G is defined by
\be
G_{\mu\sigma,\mu'\sigma'}=
\frac{\delta^{2}\!S}{\delta\!X^{\mu\sigma}\delta\!X^{\mu'\sigma'}}
- \Gamma^{\lt}_{\mu\sigma,\mu'\sigma'}\:\frac{\delta\!S}{\delta\!X^{\lt}},
\ee
and $G^{\mu\sigma,\mu'\sigma'}$ is the inverse of $G_{\mu\sigma,\mu'\sigma'}$.
The connection $\Gamma$ is introduced in order to preserve covariance under
the transformations given by eq.(4), and it will be further specified later 
on. The term $M_{\mu\sigma,\mu'\sigma'}$ is related to the
Jacobian of a change of variables, as explained in [1],
and it depends on the connection $\Gamma$ alone. In
this paper, we only need the terms linear in its expansion in terms of
 $\Gamma$:
\be
M_{\mu\sigma,\mu'\sigma'}=-\frac{1}{3}\left(\frac{\delta \Gamma^
{\lt}_{\lt,\mu\sigma}}{\delta X^{\mu'\sigma'}}+ \frac{\delta \Gamma^{\lt}_
{\lt,\mu'\sigma'}}{\delta X^{\lt}}+\frac{\delta \Gamma^{\lt}_{\mu\sigma,
\mu'\sigma'}}{\delta X^{\lt}}\right)+\cdots .
\ee
In the preceding equations, as well as in the rest of the paper, the
summation convention is also applied to the world sheet variables; repeated
variables are to be integrated over. We also frequently use the matrix(
operator) notation for expressions with two sets of indices, for example,
$G_{\mu\sigma,\mu'\sigma'}$ is to be thought of as a matrix in the set of
indices $\mu\sigma$ and $\mu'\sigma'$, with an obvious definition of the
matrix product. Another convention we follow throughout the paper is to 
write only the set equations corresponding to $v_{+}(\ssp)$, when the set
involving $v_{-}(\sm)$ can be obtained from the first set by the obvious
substitution $+ \leftrightarrow -$.  Following this convention in the
above set of equations,  we have not displayed the set corresponding
to $v_{-}$. Also, the trace in the expression $Trlog(G)$ is over the same
set of indices.

The set of eqs.(8,9,10) form the starting point of this paper;
 they are the analogue of the
renormalization group equations of reference [10].
ompared to [10], it has the advantage of being invariant under the
transformations of eq.(4), which,
 as we shall see, is important in eliminating certain spurious states. In
contrast, in the non-covariant approach of [10], there does not
seem to be enough gauge invariance to decouple all the spurious states.

As they stand, eqs.(8,9,10) are still only formal, since we have not yet
specified any cutoff or regularization procedure. We now briefly discuss
 the heat kernel method, 
 the regularization procedure we are going to use. It differs from the naive
cutoff used in [1], and it has several advantages over it:
It is simple to implement, and it preserves invariance under coordinate
transformations (eq.(4)).
There is a further advantantage in using the heat kernel method :
 Although we have written down eqs.(8,9 and 10) in full
 generality,
we are really  interested only in the local terms  in these
equations. By this, we mean terms that have a local expansion similar
to the expansion for $S^{(1)}$ in eq.(1). These terms are the only ones
 to be considered in a renormalization group analysis such as ours,
since only they contribute to the renormalization of the original local
action. The heat kernel method provides a very convenient way of extracting
these local terms, and it will enable us later on to write a finite and
 local version of the  equations (8,9,10).
 
There are two divergent terms that need regularization: The $Trlog(G)$ term in
eq.(9) and $M_{\mu\sigma,\mu'\sigma'}$ in eq.(12). Let us first consider
the $Trlog(G)$. We set
\be
2\:G_{\mu\sigma,\mu'\sigma'}=2\:\Delta_{\mu\sigma,\mu'\sigma'}+
H_{\mu\sigma,\mu'\sigma'},
\ee
where,
$$
\Delta_{\mu\sigma,\mu'\sigma'}= \eta_{\mu\mu'}\Delta_{\sigma,\sigma'},\;\;\;
\Delta_{\sigma,\sigma'}=-\pdp\dm \delta^{2}(\sigma-\sigma').
$$
We shall also need the free propagator $\Delta^{\mu\sigma,\mu'\sigma'}$,
 which is the inverse of
$\Delta_{\mu\sigma,\mu'\sigma'}$. It satisfies
$$
\Delta^{\mu\sigma,\mu'\sigma'}= \eta^{\mu\mu'} \Delta^{\sigma,\sigma'},\;\;\;
\pdp\dm\Delta^{\sigma\sigma'}=-\delta^{2}(\sigma-\sigma'),
$$
and its regularized form is given by
\be
\Delta^{\mu\sigma,\mu'\sigma'}\rightarrow
 \Delta^{\mu\sigma,\mu'\sigma'}(\epsilon)=
 \int^{\infty}_{\epsilon}dt\: \tilde{G}^{(0)}_{\mu\sigma,\mu'\sigma'}(t),
\ee
with
\bq
\tilde{G}^{(0)}_{\mu\sigma,\mu'\sigma'}(t)=
\theta(t)\left(e^{- t\Delta}\right)_{\mu\sigma,\mu'\sigma'}&=&
\eta_{\mu\mu'}\:\tilde{G}^{(0)}_{\sigma,\sigma'}(t)\nonumber\\
&=&\eta_{\mu\mu'}\:\frac{\theta(t)}{4\pi t}
 \exp\left(-\frac{(\sigma-\sigma')^{2}}{4 t}\right).
\eq
The term $Trlog(G)$ is regularized by
\be
Trlog G\rightarrow -\int^{\infty}_{\epsilon}\frac{dt}{t}\:Tr(\tilde{G}),
\ee
where the full heat kernel $\tilde{G}$ is defined by
$$
\tilde{G}_{\mu\sigma,\mu'\sigma'}(t)\equiv \theta(t)\left(e^{-t G}\right)_
{\mu\sigma,\mu'\sigma'}.
$$
We now compute the conformal variation of  $Trlog(G)$.. Starting with
\be
\delta_{v}( Trlog(G))=\int^{\infty}_{\epsilon}
dt\: Tr\left(e^{-tG}\delta_{v}( G)\right),
\ee
it is convenient to split it   into two terms:
\bq
\delta_{v_{+}}\left(G_{\mu\sigma,\mu'\sigma'}\right)&=& \partial_
{\sigma_{+}}\left(v(\ssp) G_{\mu\sigma,\mu'\sigma'}\right) +
\partial_{\sigma'_{+}}\left(v(\sigma'_{+}) G_{\mu\sigma,\mu'\sigma'}\right)
+\delta^{(2)}_{v_{+}}\left(G_{\mu\sigma,\mu'\sigma'}\right) \nonumber\\
&=& \left(\delta^{(1)}_{v_{+}}+\delta^{(2)}_{v_{+}}\right)
G_{\mu\sigma,\mu'\sigma'}.
\eq
This split is motivated by the observation that conformal transformations
are a special case of the coordinate transformations; the transformation
law of  a tensor such as
$G_{\mu\sigma,\mu'\sigma'}$ under the coordinate transformations
 contains two types of terms: the first type comes from the transformation
of the indices of the tensor and it is represented by $\delta^{(1)}_
{v_{+}}$ or the first two terms on the right hand side of eq.(18). The
second type of term corresponds to the transformation of the coordinates
on which the tensor depends and it is given by $\delta^{(2)}_{v_{+}}$.
For example, acting on the first term in eq.(11)
 for $G$, $\delta^{(2)}_{v_{+}}$ is given by
\be
\delta^{(2)}_{v_{+}}\left(\frac{\delta^{2}S}{\delta X^{\mu\sigma}
\delta X^{\mu'\sigma'}}\right)=\frac{\delta^{2}}{\delta X^{\mu\sigma}
\delta X^{\mu'\sigma'}}\left(\int d^{2}\tau\: v(\tau_{+})\:\pdp X^{\lt}
\frac{\delta S}{\delta X^{\lt}}\right).
\ee
We shall later see that all of the cutoff independent useful information
will come from $\delta^{(1)}_{v_{\pm}}$; $\delta^{(2)}_{v_{\pm}}$ will
only contribute cutoff dependent terms which will cancel.

We now turn to the evaluation of the right hand side of eq.(17).
Using the definition of the heat kernel,
 and the identity
\be
\left(v(\ssp)\partial_{\ssp}+ v(\sigma'_{+})\partial_{\sigma'_{+}}\right)
\tilde{G}^{(0)}_{\mu\sigma,\mu'\sigma'}(t)= -\frac{v(\ssp)- v(\sigma'_{+})}
{\ssp - \sigma'_{+}}\:\frac{\partial}{\partial t}\left(t\: \tilde{G}^{(0)}_
{\mu\sigma,\mu'\sigma'}(t)\right),
\ee
 one can easily
establish the following result:
\bq
&&\delta^{(1)}_{v_{+}}\left(Trlog(G)\right)= \nonumber \\
&& = \frac{1}{2}\int^{\infty}_{\epsilon} dt
\ids\idsp\:\tilde{G}_{\mu\sigma,\mu'\sigma'}\left(\partial_{\sigma_{+}}(t)
\left(v(\ssp) H_{\mu\sigma,\mu'\sigma'}\right)+ \partial_{\sigma'_{+}}
\left(v(\sigma'_{+}) H_{\mu\sigma,\mu'\sigma'}\right)\right) \nonumber \\
&& =\frac{1}{2}
 \int^{\infty}_{\epsilon}dt \ids\idsp\frac{v(\ssp)-v(\sigma'_{+})}
{\ssp -\sigma'_{+}}\bigg(H_{\mu\sigma,\mu'\sigma'}\frac{\partial}
{\partial t}\left(t \tilde{G}^{(0)}_{\mu'\sigma',\mu\sigma}(t)\right)
\nonumber \\
&&-\frac{1}{4}
 \int^{+\infty}_{-\infty}dt' \left(H\tilde{G}(t')H\right)_{\mu\sigma,
\mu'\sigma'}\frac{\partial}{\partial t}\left((t-t')\tilde{G}^{(0)}_
{\mu'\sigma',\mu\sigma}(t-t')\right)\bigg).
\eq
This equation enables us to make a clean seperation between local and
non-local contributions to eqs.(8,9,10).
We note that  the integrand is a total derivative with respect to  the variable
t.  It can therefore be integrated, with the result that the contribution
from the upper limit $\infty$ is the non-local part of the integral, and 
the contribution from the lower limit $\epsilon$ is the local part. This
follows from the well-known properties of the heat kernel, which describes
the diffusion of a point source as a function of time t. For small t,
$t=\epsilon$, the source can diffuse only a small distance in space, and so
in the limit $\epsilon\rightarrow 0$, the contribution from the lower limit
is local. On the other hand, the contribution from the upper limit is 
clearly non-local, since t is very large. We now define $\delta_{\Lambda}$
( eqs.(9,10)) so as to cancel the unwanted contribution
from the upper limit of integration over t:
\bq
\left(\delta^{(1)}_{v_{+}}+\delta_{\Lambda}\right) Trlog(G)& =&
-\frac{1}{2}\ids\idsp \vessup
\bigg(\epsilon\:\tilde{G}^{(0)}_
{\mu'\sigma',\mu\sigma}
(\epsilon) H_{\mu\sigma,\mu'\sigma'} \nonumber\\
&-& \frac{1}{2}\int^{\epsilon}_{0} dt'(\epsilon -t') \tilde{G}^
{(0)}_{\mu'\sigma',\mu\sigma}(\epsilon -t') \left(H \tilde{G}H\right)_
{\mu\sigma,\mu'\sigma'}\bigg).
\eq
It can easily be shown that this definition ensures that the long distance
behavior of the free propagator is unchanged under conformal transformations.
For this reason, in the case of free propagator, the regularization we
are using agrees with
 the cutoff used in reference [10]. Having extracted the local
part of eq.(9), we can rewrite eqs.(9,10) in the following form:
\bq
&&b\ids \idsp\: v(\sigma'_{+}) F^{\mu\sigma}_{\sigma'}(X)\: \frac{\delta S}
{\delta X^{\mu\sigma}}+ \frac{1}{2}\ids\idsp \vessup \nonumber \\
&&\times \left(\frac{1}{2} \epsilon\:
\tilde{G}^{(0)}_{\mu'\sigma',\mu\sigma}(\epsilon) H_{\mu\sigma,
\mu'\sigma'}-\frac{1}{4}
 \int dt' (\epsilon - t') \tilde{G}^{(0)}_{\mu'\sigma',\mu\sigma}
(\epsilon - t')\left(H \tilde{G}(t') H\right)_
{\mu\sigma,\mu'\sigma'}\right) \nonumber \\
&&- \frac{1}{2}\int^{\infty}_{\epsilon} dt\:
 \tilde{G}_{\mu'\sigma',\mu\sigma}(t) \bigg( 
 \int d^{2}\tau \int d^{2}\tau'\: v(\tau'_{+}) f^{\lt}_{\tau'}(X)
\:\frac{\delta G_{\mu\sigma,\mu'\sigma'}}{\delta X^{\lt}}\nonumber\\
&&+\frac{1}{2}\:\delta^{(2)}_{v}(H_{\mu\sigma,\mu'\sigma'})\bigg)+E_{M}=0.
\eq
We shall often need the part of the above equation linear in fields.
It is quite straightforward to linearize various terms except perhaps $E_{M}$.
 Since
$M$ already starts at the linear order(eq.(12)), in the factor in front of
 this term, we can replace $G$  by the zeroth order term in its expansion:
$$
G^{\mu\sigma,\mu'\sigma'}\rightarrow \int^{\infty}_{\epsilon} dt\:
\tilde{G}^{(0)}_{\mu\sigma,\mu'\sigma'}(t),
$$
and arrive at the result
\bq
&&- F^{\lt}_{v}\: \frac{\delta G^{\mu\sigma,\mu'\sigma'}}{\delta X^{\lt}}
+ \frac{\delta F^{\mu\sigma}_{v}}{\delta X^{\lt}}\: G^{\lt,\mu'\sigma'}
+ \frac{\delta F^{\mu'\sigma'}_{v}}{\delta X^{\lt}}\: G^{\mu\sigma,\lt}
- \delta_{\Lambda}(G^{\mu\sigma,\mu'\sigma'}) \nonumber \\
&&\rightarrow \int^{\infty}_{\epsilon} dt \left(v(\ssp) \partial_{\ssp}
+ v(\sigma'_{+})\partial_{\sigma'_{+}}- \delta_{\Lambda}\right)
\tilde{G}^{(0)}_{\mu\sigma,\mu'\sigma'}(t) \nonumber \\
&&=  \vessup\: \epsilon\: \tilde{G}^{(0)}_{\mu\sigma,\mu'\sigma'}
(\epsilon),
\eq
Next, we turn our attention to eq.(12) for $M_{\mu\sigma,\mu'\sigma'}$.
This expression needs regularization, since the integration over the variable
$\tau$ will lead to divergences. Again, we use the heat kernel method to
regulate it.  Making use of a basic property of the heat kernel,namely, as
$\epsilon\rightarrow 0$,
$$
lim\left(\tilde{G}^{(0)}_{\mu\sigma,\mu'\sigma'}(\epsilon)\right)\rightarrow
\eta_{\mu\mu'}\: \delta^{2}(\sigma - \sigma'),
$$
we can regularize the integration over $\tau$ by setting, for example
$$
\int d^{2}\tau\: \Gamma^{\lt}_{\lt,\mu\sigma}\rightarrow \int d^{2}\tau
\int d^{2}\tau'\: \Gamma^{\lt}_{\lambda'\tau',\mu\sigma} \tilde{G}^{(0)}_
{\lt,\lambda'\tau'}(\epsilon).
$$
Combining this with eq.(24) yields the following regulated expression for
$E_{M}$:
 \bq
E_{M}&=& -\frac{1}{3}\ids\idsp \int d^{2}\tau \int d^{2}\tau'
\vessup \epsilon\:\tilde{G}^{(0)}_{\mu\sigma,\mu'\sigma'}(\epsilon) 
\tilde{G}^{(0)}_{\lt,\lambda'\tau'}(\epsilon) \nonumber \\
&\times & \left(2\: \frac{\delta \Gamma^{\lt}_{\lambda'\tau',\mu\sigma}}
{\delta X^{\mu'\sigma'}}+ \frac{\delta \Gamma^{\lt}_{\mu\sigma,
\mu'\sigma'}}{\delta X^{\lambda'\tau'}}\right)+ \cdots,
\eq
where the dots represent higher order terms in the fields that we have
not written down. Finally, putting everything together, we have the
following linear version of eq.(23):
\bq
&&b\ids\:  v(\sigma_{+})\: \pdp X^{\mu\sigma}\frac{\delta S^{(1)}}{\delta
X^{\mu\sigma}}
-  2b \ids\idsp v(\sigma'_{+})\: f^{\mu\sigma}_{\sigma'}(X)\;
\pdp\dm (X^{\mu\sigma}) \nonumber \\
&& +\frac{1}{4}\ids\idsp \vessup\: \epsilon\:
 \tilde{G}^{(0)}_{\mu'\sigma',
\mu\sigma}(\epsilon) H_{\mu\sigma,\mu'\sigma'}\nonumber \\
&&-\frac{1}{4} \int^{\infty}_{\epsilon}
dt\: \tilde{G}^{(0)}_{\mu'\sigma',\mu\sigma}(t)\: \delta^{(2)}_{v}
(H_{\mu\sigma,\mu'\sigma'}) + E_{M}=0,
\eq
where,
\bq
H_{\mu\sigma,\mu'\sigma'}&=& \frac{\delta^{2}S^{(1)}}{\delta X^
{\mu\sigma}\delta X^{\mu'\sigma'}}- \Gamma^{\lt}_{\mu\sigma,\mu'\sigma'}
\frac{\delta S}{\delta X^{\lt}}\nonumber\\
&\simeq& \frac{\delta^{2}S^{(1)}}{\delta X^{\mu\sigma}\delta X^{\mu'\sigma'}}
+2\:\Gamma^{\lt}_{\mu\sigma,\mu'\sigma'}\pdp\dm X^{\lt}.
\eq

In closing this section, let us comment on what has been accomplished
so far. Using the heat kernel method, we have both regularized the basic
renormalization group equations (8,9,10) and also extracted their
local component. The result is eq.(23) and its linearized version, eq.(26).
At first sight, it is not clear that these equations are powerful
enough to give useful information.
 For example,
 the connection 
$\Gamma$ and the field $f$ (eq.(7)) were introduced
 as independent fields in our
equations. On the other hand, since the string field equations
 should ultimately be
expressible only in terms of the string fields that appear
 in the basic action of eq.(1),
 $\Gamma$ and $f$ should somehow be eliminated
in favor of these fundamental fields.
  In the following sections, we shall see that there is no need of
 an a priori
determination of the auxilliary fields
 $\Gamma$ and $F$ ; the equations
themselves will do this job for us. The situation is somewhat similar to the
first order formulation of general relativity, when the connection is
introduced as an independent field and then determined from the equations
of motion. We have somewhat oversimplified the situation here; the equations
we have, unlike those in general relativity,
are not quite powerful enough to determine all the components of these
auxiliary fields. However, the undetermined components are also
unneeded; they do not appear in the  equations for the string fields.
 As a consequence,
the auxiliary fields can be completely eliminated from the final string
field equations.

Another question concerns the cutoff dependence of the equations. We shall
see that the equations neatly seperate into cutoff independent and 
cutoff dependent parts. The cutoff dependent pieces have a different
structure than the cutoff independent ones, and as a result, they have to
cancel among themselves. The resulting equations partially
 fix $\Gamma$ and $F$, but they do not lead to any relations between
 the string fields. As we shall see, the useful equations come
exclusively from the cutoff independent pieces in eq.(26).

In the next two sections, eq.(26) will be applied to the massless and
the first massive levels of the string, neglecting all the rest of the 
levels. In section 5, we generalize our treatment to include all of the
 levels. The reasons for specializing to these two levels are the following:
In [1], we considered the same two levels
of the string, with somewhat unsatisfactory results for the
first massive level. We feel that it is instructive to compare the improved
treatment given here to the treatment given in [1], and to
show that all the dificulties encountered in the earlier paper are easily
overcome. In addition, since the general treatment of all the levels given 
in section 5 is somewhat formal, we felt that working out two simple
examples in some detail might be useful.

\vskip 9pt
\noindent{\bf 3. Linearized Equations For the Zero Mass States}
\vskip 9pt

In this section, we apply eq.(26) to the massless states of the string, the
graviton, the dilaton and the antisymmetric tensor, suppressing for the
time being all the other states of the string. Also, we confine ourselves
 to a linearized
treatment, which serves as an introduction to the full 
non-linear treatment of section 6. The action that describes the graviton 
and the antisymmetric tensor is the second term in eq.(1):
\be
S^{(1)}=\ids\:\tilde{h}_{\mu\nu}\left(X(\sigma)\right)\pdp X^{\mu\sigma}
\dm X^{\nu\sigma}.
\ee
The dilaton is at the moment missing, and it will make its appearence
later as part of  the connection. As explained earlier, the connection 
will not be specified yet, and only the following general conditions will
be imposed on it:\\
a) The connection should be a local function of $ X(\sigma)$.\\
b) Its classical conformal dimension should be determined by requiring that
 the two terms in eq.(11) have the same dimension. 
 This requirement guarantees 
that $G_{\mu\sigma,\mu'\sigma'}$ will have a well defined classical conformal
 dimension.\\
These requirements fix the form of $\Gamma$ to be
\be
\Gamma^{\lt}_{\mu\sigma,\mu'\sigma'}= \Gamma^{\lambda}_{\mu\mu'}
\left(X(\sigma)\right) \delta^{2}(\tau - \sigma) \delta^{2}(\sigma
-\sigma').
\ee
We should make it clear that although we are using for it the same symbol
as the usual metric derived connection of general relativity,
 $\Gamma^{\lambda}_{\mu\mu'}$ is as yet an undetermined function
of $X(\sigma)$. In fact, in the end, it will turn out to be different
from the standard result.

We now discuss the expected invariances of the model. Since all
the higher levels in the action are neglected, invariance under coordinate
transformations, eq.(4),  is restricted to the coordinate transformations
of general relativity,
\be
X^{\mu\sigma}\rightarrow X^{\mu\sigma}+ f^{\mu}\left(X(\sigma)\right).
\ee
 In addition to these coordinate transformations,
there is invariance under the gauge transformations given eq.(3). Taking
$$
I_{+}=-\pdp X^{\mu\sigma} \Lambda_{\mu},\;\;\;
I_{-}=\dm X^{\mu\sigma} \Lambda_{\mu},
$$
we have the well-known gauge transformations of the antisymmetric tensor:
\be
B_{\mu\nu}\rightarrow B_{\mu\nu}+\partial_{\mu}\Lambda_{\nu}
- \partial_{\nu}\Lambda_{\mu}.
\ee 
Finally, the action given by eq.(28) is conformally invariant in the classical
limit.

 As a consequence of these invariances, in computing the contribution of
various terms to eq.(26),
a number of simplifications occur:\\
a) Two of the terms in (26) vanish as a result of the conformal invariance
of the action,
\bq
\ids\: v(\ssp)\: \pdp X^{\mu\sigma}\:\frac{\delta S^{(1)}}{\delta X^
{\mu\sigma}}&=&0,\nonumber\\
\delta^{(2)}_{v}(H_{\mu\sigma,\mu'\sigma'})&=&0. 
\eq
b) The second term in the same equation also vanishes, since $f^{\mu\sigma}_
{\sigma'}=0$. This is because the first term for $F$ in eq.(7) already
transforms as a vector under (30).\\
The remaining of the terms are given by,
\bq
&&\ids\idsp \vessup\epsilon\:\tilde{G}_{\mu'\sigma',\mu\sigma}(\epsilon)
H_{\mu\sigma,\mu'\sigma'} \nonumber \\
&& =\frac{1}{4\pi}\ids\: v'(\ssp) \pdp X^{\mu\sigma}\dm X^{\mu'\sigma}
\Big(\Box\tilde{h}_{\mu\mu'}-
\partial_{\mu}\partial_{\lambda}\tilde{h}_{\lambda\mu'} \nonumber \\
&&+\partial_{\mu'}\partial_{\lambda}\tilde{h}_{\lambda\mu}-2\:
\partial_{\mu'}\Gamma^{\mu}_{\lambda\lambda}\Big) + \frac{1}{2\pi\epsilon}
\ids\: v'(\ssp)\tilde{h}_{\lambda\lambda},\\
&&E_{M}=-\frac{1}{3(4\pi)^{2}\epsilon}\ids\: v'(\ssp)\left(2\partial_{\mu}
\Gamma^{\nu}_{\nu\mu}(X(\sigma))+\partial_{\nu}
\Gamma^{\nu}_{\mu\mu}(X(\sigma))\right).
\eq
Substituting these results in eq.(26), we note that terms propotional to
the factor $\pdp X^{\mu\sigma} \dm X^{\mu'\sigma}$ and terms that do not 
have this factor must cancel seperately among themselves. Since the terms
 without this factor are proportional to $1/\epsilon$, it follows that
cutoff dependent terms cancel among themselves, and the cutoff factor does not
appear in the resulting equations:
\bq
& & \Box\tilde{h}_{\mu\mu'}-
\partial_{\lambda}\partial_{\mu}\tilde{h}_{\lambda\mu'}+\partial_{\mu'}
\partial_{\lambda}\tilde{h}_{\lambda\mu}-2\:\partial_{\mu'}\Gamma^{\mu}_
{\lambda\lambda}=0, \\
& &\tilde{h}_{\lambda\lambda}- \frac{1}{6\pi}(2\:\partial_{\mu}\Gamma^
{\lambda}_{\lambda\mu}+\partial_{\lambda}\Gamma^{\lambda}_{\mu\mu})
=0.
\eq
In addition to eq.(34), which came from conformal transformations on the
variable $\ssp$, there is a $\sm$ counterpart, obtained by letting
$$
\tilde{h}_{\mu\mu'}\rightarrow \tilde{h}_{\mu'\mu}
$$
in that equation:
\be
\Box \tilde{h}_{\mu'\mu}-\partial_{\mu}\partial_{\lambda}\tilde{h}_
{\mu'\lambda}+\partial_{\mu'}\partial_{\lambda}\tilde{h}_{\mu\lambda}
-2\:\partial_{\mu'}\Gamma^{\mu}_{\lambda\lambda}=0.
\ee
Combining eqs.(34) and (36) yields the following result:
\bq
& &\Gamma^{\lambda}_{\mu\mu}=\partial_{\mu} h_{\mu\lambda}-\frac{1}{2}
\partial_{\lambda}h_{\mu\mu}
+\partial_{\lambda}
\phi, \\
& &\Box h_{\mu\mu'}-\partial_{\mu}\partial_{\lambda}h_{\mu'\lambda}-
\partial_{\mu'}\partial_{\lambda}h_{\mu\lambda}+\partial_{\mu}
\partial_{\mu'}h_{\lambda\lambda}
-2\:\partial_{\mu}
\partial_{\mu'}\phi=0, \\
& &\Box B_{\mu\mu'}-\partial_{\mu}\partial_{\lambda} B_{\lambda\mu'}
+ \partial_{\mu'}\partial_{\lambda}B_{\lambda\mu}=0.
\eq
We now make a few observations:\\
a) Eq.(38) determines only $\Gamma^{\lambda}_{\mu\mu}$, the contracted
part of the connection, up to a total derivative  of a
new field. We identify this field $\phi$ with the dilaton field.\\
b) Eq.(39) is the correct linearized equation for the gravitational
 field (symmetric
part of $\tilde{h}$), coupled to the dilaton field. \\
c) Eq.(40) is the correct linearized equation for the antisymmetric tensor.\\
d) Those components of $\Gamma$ not determined by eqs.(35) and (36) play no
 role in the equations for the fundamental string fields. In fact, $\Gamma$
is completely absent from the equations for $h$ and $B$. One can think of
this as some kind of gauge invariance operating on $\Gamma$, although we
will not stress this point of view in this paper. \\
e) The linearized equation for the dilaton
$$
\Box \phi=0,
$$
is still missing. It can in fact be derived from the eq.(38) for gravity
as follows: This equation is invariant under 
\be
h_{\mu\nu}\rightarrow h_{\mu\nu}+\partial_{\mu}\kappa_{\nu}+
\partial_{\nu}\kappa_{\mu},
\ee
the standard linearized gauge transformations of gravity. If the 
d'Alembertian acting on $h_{\mu\mu'}$ in (39) is invertible, then $h$ is
a pure gauge. Therefore, the only physical part of $h$ comes from the
non-invertible part of the d'Alembertian. One can then fix the gauge
so that
$$
\Box h_{\mu\nu}=0.
$$
In this gauge, applying the d'Alembertian on both sides of (38), we find
that $\phi$ satisfies the massles free field equation.

To summarize, we have shown in this section that eq.(26) correctly
reproduces the coupled gravity-dilaton equations in the linear approximation.
We stress that no a priori choice of metric or connection was made; in
fact, the metric played no role at all in our derivation. This is in
contrast to the standard treatment [2-6], where an initial choice
of the metric is made. The problem  is that when
the higher levels of the string are present, they will in general
contribute to the metric, it is no longer easy to guess the form of
this contribution.
 An incorrect initial choice would in general conflict with the
renormalization group equations. We avoid this problem by letting the
equations determine as much of the connection as possible;
 the components of the connection that
are left undetermined are spurious and do not appear in the equation for the
physical fields.
 The dilaton
field, which was not present in the original action(eq.(1)), emerges from
 as part of the connection. Again, 
this differs from the standard approach [5],
which introduces the dilaton field  in the original action.

\vskip 9pt
\noindent {\bf 4. Linearized Equations For The First Massive Level}
\vskip 9pt

In this section, we apply eq.(26) to the first massive level of the string.
In general, the action for the first massive level contains 8 terms; however,
using gauge transformations of the form given by eq.(3), it was shown in [1]
 that five of those terms can be eliminated, resulting in the
following completely gauge fixed form:
\bq
S^{(1)}&=&\ids \Big(e_{\mu_{1}\mu_{2},\nu_{1}\nu_{2}}\pdp X^{\mu_{1}}
\pdp X^{\mu_{2}}\dm X^{\nu_{1}}\dm X^{\nu_{2}} \nonumber \\
&+& e_{\mu_{1}\mu_{2}\mu_{3}}\pdp\dm X^{\mu_{1}}\pdp X^{\mu_{2}}
\dm X^{\mu_{3}} +e_{\mu_{1}\mu_{2}} \pdp\dm X^{\mu_{1}}\pdp\dm X^{\mu_{2}}
\Big).
\eq
In this formula, the fields, as usual, are assumed to be local functions
of $X(\sigma)$. The full action is again sum of free and interacting terms:
$$
S=S^{(0)}+S^{(1)},
$$
with $S^{(0)}$ given by eq.(1). The coordinate transformations relevant
for this action are
\be
X^{\mu\sigma}\rightarrow X^{\mu\sigma}+ f^{\mu}_{\nu\lambda}(X(\sigma))
\pdp X^{\nu\sigma}\dm X^{\lambda\sigma}+ f^{\mu}_{\nu}(X(\sigma))
\pdp\dm X^{\nu\sigma}.
\ee
These transformations, acting on $S^{(0)}$, generate terms of the same
form as the terms proportional to $e_{\mu_{1}\mu_{2}\mu_{3}}$ and
$e_{\mu_{1}\mu_{2}}$ in $S^{(1)}$. In fact, these terms can be eliminated
by choosing
\be
2\:f^{\mu_{1}}_{\mu_{2}\mu_{3}}= e_{\mu_{1}\mu_{2}\mu_{3}},\;\;\;
f^{\mu_{1}}_{\mu_{2}}+f^{\mu_{2}}_{\mu_{1}}= e_{\mu_{1}\mu{2}},
\ee
which shows that the corresponding states are spurious, so long as the
theory is invariant under (43). The only set of states which cannot be
 decoupled are the states represented by the first term in eq.(42); these
are therefore the only physical states.

\noindent At this
point, we would like to make the following observations:\\
a) The coordinate transformations , acting $S{(1)}$, generate additional
terms. Since we are investigating only the linear portion of the theory,
these terms can be neglected.\\
b) The terms eliminated by coordinate invariance are the same terms which
would vanish, if the free equations of motion for X,
\be
\pdp\dm X^{\mu\sigma}=0,
\ee
were imposed. Of course, in the linearized theory, the free field equations are
what remain from the full set of interacting classical equations that
 follow from
the action (1). Covariantizing the theory with respect to coordinate
transformations therefore enables one to use the
classical equations of motion in conjunction with
the renormalization group equations. We should stress that, since the
renormalization group equations deal with off mass shell quantities,
the use of the classical equations of motion is in general not
 permissible in a non-covariant approach. This is clearly a good feature
of the covariant approach, since the states eliminated by the equations
of motion are also absent in the
standard treatment of the string theory. In contrast, in a non-covariant
treatment of the renormalization group equations, it is not clear how to
eliminate these unwanted states [10].
It is also interesting to know whether what we are doing here is  related
 to the Batalin-Vilkovisky program [21,22],
 which also makes it possible
to use the equations of motion in an off-shell formulation. In this context,
Henneaux [23] discussed the connection between field redefinitions,
equations of motion and Batalin-Vilkovisky method.\\
c) Let us compare the coordinate transformations of
general relativity(eq.(30)), which are completely local on the world sheet,
with the transformations of eq.(43), which, in contrast,
 contain derivatives with respect
to the world sheet coordinates. Invariance under either set of transformations
serves to eliminate spurious states. There is, however, a difference:
Invariance under diffeomorphisms of general relativity, in contrast
 to invariance
under (43),
 does not lead
to  on mass shell constraints .\\ 

The next step is to secure invariance under (43) by introducing a suitable
connection. The conditions that the connection must satisfy are the same
ones that lead to eq.(29); namely, locality and the correct classical
conformal dimension, plus world sheet Lorentz invariance. The expansion in
terms of delta functions in world sheet coordinates and their derivatives
is rather lengthy; it contains ten terms. To give the reader an idea, we
exhibit a few typical terms below:
\bq
\Gamma^{\lt}_{\mu\sigma,\mu'\sigma'}&=&\delta^{2}(\tau-\sigma)
\delta^{2}(\sigma-\sigma')\Big(\Gamma^{(1)\lambda}_{\mu\mu',
\alpha\beta}(X(\sigma))\: \pdp X^{\alpha\sigma}\dm X^{\beta\sigma}\nonumber\\
&+&\Gamma^{(2)\lambda}_{\mu\mu',\alpha}\:\pdp\dm X^{\alpha\sigma}\Big)+
\delta^{2}(\tau -\sigma)\partial_{\ssp}\delta^{2}(\sigma -\sigma')
\Gamma^{(3)\lambda}_{\mu\mu',\alpha}\: \dm X^{\alpha\sigma}\nonumber\\
&+&\delta^{2}(\tau -\sigma) \dsm \delta^{2}(\sigma -\sigma')\Gamma^
{(4)\lambda}_{\mu\mu',\alpha}\: \pdp X^{\alpha\sigma}+ \cdots
\eq
The reader should have no trouble in constructing the remaining terms acording 
to the following rules: There are always two delta functions setting the
worldsheet variables $\sigma$,$\sigma'$ and $\tau$ equal to each other,
and there is one derivative with respect to $\ssp$ and one derivative
with respect $\sm$, acting on the delta functions or on $X$'s. Each term
contains also a local function of $X(\sigma)$, denoted by $\Gamma$'s with
superscripts. 
	In a similar fashion, using locality and dimensional
analysis,
 the unknown function in the definition of the conformal
Killing vector (eq.(7)) can be written as
\be
\int d^{2}\tau \:v(\tau_{+}) f^{\mu\sigma}_{\tau}= v'(\ssp)\left(
f^{(1)\mu}_{\nu\lambda}\:\pdp X^{\nu\sigma} \dm X^{\lambda\sigma}
+ f^{(2)\mu}_{\nu}\:\pdp\dm X^{\nu\sigma}\right).
\ee
We now substitute eqs.(42),(46) and (47) into (26); the resulting equations
are the linear part of the string field equations satisfied by the states
  at the first massive level. 	Since
 these equations are rather lengthy and
a knowledge of their detailed form is not particularly important, in what
 follows  some their important general features will be described, and
 a few of them that are really needed will be written down. First of all,
it is useful to exhibit the cutoff dependence of the equations 
 by writing them in the following form:
\bq
&&\ids\: v'(\ssp)\Big( \pdp X^{\mu_{1}\sigma}\pdp X^{\mu_{2}\sigma}
\dm X^{\nu_{1}\sigma}
\dm X^{\nu_{2}\sigma}
A^{(1)}_{\mu_{1}\mu_{2},\nu_{1}\nu_{2}}(X(\sigma))\nonumber\\
&&+\pdp\dm X^{\mu_{1}}\pdp X^{\mu_{2}}\dm X^{\mu_{3}} A^{(2)}_{\mu_{1}
\mu_{2}\mu_{3}}
 +\pdp\dm X^{\mu_{1}}\pdp\dm X^{\mu_{2}}A^{(3)}_{\mu_{1}\mu_{2}}\nonumber\\
&&+\partial^{2}_{+} X^{\mu_{1}}\dm X^{\mu_{2}}\dm X^{\mu_{3}} A^{(4)}_
{\mu_{1}\mu_{2}\mu_{3}}+ \partial^{2}_{+}X^{\mu_{1}}\dm X^{\mu_{2}}
A^{(5)}_{\mu_{1}\mu_{2}}\Big)\nonumber\\
&&+ log(\epsilon)\ids\:v'(\ssp)
\left(\pdp X^{\mu_{1}\sigma}\pdp X^{\mu_{2}\sigma}
\dm X^{\nu_{1}\sigma}\dm X^{\nu_{2}\sigma}B^{(1)}_{\mu_{1}\mu_{2},\nu_{1}
\nu_{2}}+\cdots \right)\nonumber\\
&&+\frac{1}{\epsilon}\ids\:v'(\ssp) \pdp X^{\mu\sigma} \pdp X^{\nu\sigma}
C_{\mu\nu}(X(\sigma)) + \frac{1}{\epsilon^{2}}\ids\:v'(\ssp)
 D(X(\sigma))\nonumber\\
 &&=0.
\eq
Without doing any calculation, the general form given above follows again
from locality and naive dimensional analysis. 
The dots stand for terms  not written down, which can be obtained
by replacing $A^{(i)}$'s by $B^{(i)}$'s in the line above them.
 We now examine each of these terms in turn:\\
a) The last two terms have a dependence on derivatives of $X$ quite
 different from the first two terms and also from each other. As a
consequence, it follows that C and D must vanish seperately,
$$
C_{\mu\nu}=0,\;\;\; D=0.
$$
These equations constrain various pieces of the connection; since they are
rather lengthy and they do not contribute to string field equations, which are
our main interest, we are not going to write them out explicitly. The
important point is that
 cutoff dependences of the form $1/\epsilon$ and $1/(\epsilon)^{2}$ have
completely dissappeared from the equations.\\
b) There is still a cutoff dependence of the form $\log(\epsilon)$ in the
second term, and since this term has exactly the same structure as the
first (cutoff independent) term, we cannot demand that it vanishes seperately.
However, after some manipulation of the equations, it is not difficult to
show that
\be
\ids\left(\pdp X^{\mu_{1}\sigma}\pdp X^{\mu_{2}\sigma}
\dm X^{\nu_{1}\sigma}\dm X^{\nu_{2}\sigma} B^{(1)}_
{\mu_{1}\mu_{2},\nu_{1}\nu_{2}}
+\cdots \right) =\frac{1}{16 \pi}S^{(1)},
\ee
where $S^{(1)}$ is given by eq.(42). Since $S^{(1)}$ is multiplied by the
slope parameter b (see eq.(9)), we can get rid of the logarithmic cutoff
dependence by redefining b:
\be
b'= b+\frac{\log(\epsilon)}{16\pi}
\ee
Exactly the same redefinition also eliminates the $\log(\epsilon)$
dependence from the equations for the tachyon and for the higher massive
levels. Therefore, the renormalization of the slope parameter, which gets
rid of terms proportional to $\log(\epsilon)$
 is the only
renormalization needed ; all the cutoff dependence drops out of the equations
automatically.\\
c) Having disposed of all the cutoff dependent
 terms in (48), we are left with five cutoff
independent equations
$$
A^{(i)}_{\mu_{1}\mu_{2},\nu_{1}\nu_{2}}=0,\;\;\; i=1,2,..,5.
$$
Three of these equations, those involving $A^{(5)}$, $A^{(2)}$ and $A^{(3)}$,
 provide further constraints on the connection
and the conformal Killing vector, whereas the remaining two, involving
$A^{(1)}$ and $A^{(4)}$, can be written exclusively in terms of the string
field $e_{\mu_{1}\mu_{2},\nu_{1}\nu_{2}}$. As the discussion leading to eq.
(44) shows, among the fields of the first massive level given by eq.(42),
 this field is the only  physical one, so we expect that the string field
equations should finally be expressible in terms of only this field.
 Taking into account the $\sm$
counterpart of (48) and simplifying, we have,
\be
\Box e_{\mu_{1}\mu_{2},\nu_{1}\nu_{2}}+16\pi b'\: e_{\mu_{1}\mu_{2},
\nu_{1}\nu_{2}}=0,
\ee
and,
\bq
& &\partial_{\nu} e_{\mu_{1}\mu_{2},\nu\nu_{1}}-\frac{1}{6}\;
 \partial_{\nu_{1}}e_{\mu_{1}\mu_{2},\nu\nu}=0,\nonumber\\
& &\partial_{\mu} e_{\mu\mu_{1},\nu_{1}\nu_{2}}-\frac{1}{6}\:
\partial_{\mu_{1}} e_{\mu\mu,\nu_{1}\nu_{2}}=0.
\eq
The above  are indeed the correct equations for the first massive level of
the closed string. In case of eq.(49), this is obvious; on the other hand,
eqs.(52) may not look familiar. This is because,
 in writing down eq.(42), we have made use of linear gauge transformations
of the form of eq.(3) to eliminate some spurious states. In the next section,
we will show that, in the string language, these correspond to gauges
generated by the operators $L_{-1}$ and $\bar{L}_{-1}$. What we have done
amounts to explicitly solving the string equations
$$
L_{1}|s>=\bar{L}_{1}|s>=0.
$$
Eqs.(50) are then equivalent to the remaining string equations
$$
L_{2}|s>=\bar{L}_{2}|s>=0.
$$
In the standard string approach, one starts with the redundant set of fields
$E_{\mu_{1}\mu_{2},\nu_{1}\nu_{2}}$, $E_{\mu,\nu_{1}\nu_{2}}$, 
$E_{\mu_{1}\mu_{2},\nu}$ and $E_{\mu\nu}$ (see Appendix B of [1])
 without initially imposing any string equations.
 The connection between our $e_{\mu_{1}\mu_{2},\nu_{1}\nu_{2}}$ and the
string field $E_{\mu_{1}\mu_{2},\nu_{1}\nu_{2}}$ is
\be
e_{\mu_{1}\mu_{2},\nu_{1}\nu_{2}}= E_{\mu_{1}\mu_{2},\nu_{1}\nu_{2}}+
\frac{5}{12}\left(\eta_{\mu_{1}\mu_{2}} E_{\mu\mu,\nu_{1}\nu_{2}}+
\eta_{\nu_{1}\nu_{2}} E_{\mu_{1}\mu_{2},\nu\nu}\right)+\frac{5}{48}
\eta_{\mu_{1}\mu_{2}}\eta_{\nu_{1}\nu_{2}} E_{\mu\mu,\nu\nu}.
\ee
It is then not difficult to show that the free string equations for the
$E$'s are equivalent to eqs.(52).

At this point, we would like to compare the results obtained here for the
first massive level to the results of [1].
 The main difference is that in [1], only
left-right symmetric models could be treated, whereas here there is no such
restriction. Also, the restriction that the coordinate transformations of
eq.(4) should have unit determinant has been removed. These restrictions
were due to an improper choice of the connection in [1];
they dissappear when the connection is freed from any a priori
constraint.

Finally, in closing this section, let us try to understand how starting
with a non-renormalizable action (eq.(42)) and a largely arbitrary
connection (eq.(46)), we were able to derive unique and cutoff independent
equations. This  result follows from the structure of eqs.(48). Since
the theory is non-renormalizable, there is a  singular dependendence on 
the cutoff, in the form of terms proportional to $1/\epsilon$ and
$1/\epsilon^{2}$. Because of their different structure, however,
 these terms satisfy
seperate equations and never mix with finite terms or terms proportional
to $log(\epsilon)$. Dimensional analysis dictates the structure of these
terms; each
additional power of $\epsilon$ must go with an additional derivative
with respect to the world sheet coordinate.
 Moreover, the equations proportional to $1/\epsilon$ and $1/\epsilon^{2}$
 determine partially
only the connection and the conformal Killing vector; they impose no
constraints on the string fields. The remaining equations are similar
those coming from a renormalizable theory; the $log(\epsilon)$ is absorbed
into slope renormalization, and at the end, one is left with the cutoff
independent equations of the form
\be
A^{(i)}_{\mu_{1}\mu_{2},\nu_{1}\nu_{2}}=0.
\ee
Furthermore, these equations can be neatly seperated into two sets:
Those that receive contribution from the connection and the Killing vector
and those that do not. The structures of these two sets are different;
the first set of terms can be written in the form
\be
\sum_{m,n}\int d^{2}\tau \:Q^{m,n}_{\lt}
\:\partial^{m}_{+}\partial^{n}_{-}X^{\lt},
\ee
where both m and n are integers $\geq 1$. The second set consists of terms
that cannot be written in this form. Or, stated otherwise,
  the first set  of terms vanish upon imposing the free field equations
(45), whereas the second set of terms do not. Glancing at eq.(48), we see that
 $A^{(2)}$, $A^{(3)}$ and $A^{(5)}$  belong to the first set and therefore
only they receive contributions from the connection and
the conformal Killing vector. On the other hand,
 $A^{(1)}$ and $A^{(4)}$ belong to the second set, and setting them equal
to zero yields the string field equations (51, 52). It is not too difficult
to see why  the connection and the Killing vector contribute only
 to the first
set: Both contributions include a factor
$$
\frac{\delta S^{(0)}}{\delta X^{\mu\sigma}}=-2\:\pdp\dm X^{\mu\sigma},
$$
which vanishes when the free field equations (45) are used.
 In the next section, we shall
see that the same seperation into two sets
 also works in the case of all the higher levels.

\vskip 9pt
\noindent {\bf 5. Linear Equations For All Levels}
\vskip 9pt

In this section, we shall show that, the standard free string equations
\bq
(L_{0}-1)|s>&=& 0,\;\;\; (\bar{L}_{0}-1)|s>= 0,\nonumber\\
L_{n}|s>&=& 0,\;\;\; \bar{L}_{n}|s>= 0,
\eq
can be derived from eq.(26). In the previous two sections, we have already
shown this for the massless and the first massive levels. Here, we present a
 general proof that applies to all the levels. In constructing the proof,
we will make use of the following results of the last section:\\
a) Only the cutoff independent part of eq.(26) gives useful information
about the string states; the cutoff dependent equations proportional
to inverse powers of $\epsilon$ provide only constraints 
on the connection and the Killing vector. This result can be established
for the higher levels without much trouble by appropriately generalizing
eqs.(46),(47) and (48) to these states. Just as in the case of the first 
massive level, the number of constraints on the connection, for example,
are far fewer than the number of allowed components of the connection,
and therefore, the connection is only partially fixed.\\
b) The terms that depend logarithmically on $\epsilon$
 turn out to be proportional to the action,
 with the same constant of proportionality as in eq.(49).
They are eliminated by slope renormalization.\\
 c)  The cutoff independent equations can be split into the two sets
discussed at the end of the last section (see eq.(55)). Only the second
set of equations, which do not vanish upon imposing the free field
equations (45), are free of the connection and the Killing vector and
hence lead to useful string field equations. This is established by 
the same argument given at the end of the last section.\\
 Let us now extract these ``useful'' equations from (26), by first
eliminating all of the  cutoff dependent
 terms and terms belonging to the first set. The third and the last terms
on the left hand side of this equation are purely cutoff dependent and can be
 dropped. One can also drop the terms proportional to the connection and
the Killing vector, since they belong to the first set.
 The only remaining term with $\epsilon$ dependence is
 the second term,
 which, in addition to
singular pieces, contains a cutoff independent subterm.
 To extract it, we note that, because of locality,
 $H$ will turn out to be the sum of terms proportional to
$\delta^{2}(\sigma-\sigma')$ or some order derivative of it with
respect to $\ssp$ and $\sm$. The terms in $H$ that are
proportional to the delta function without
the derivatives have  cutoff independendent contributions. This follows
from eq.(15) for $\tilde{G}^{(0)}$:
$$
\epsilon\:\tilde{G}^{(0)}_{\mu'\sigma',\mu\sigma}(\epsilon)
\delta^{2}(\sigma-\sigma')=
 \frac{1}{4\pi}\: \eta_{\mu\mu'}\:\delta^{2}(\sigma-\sigma').
$$
For this term, we can replace $\tilde{G}^{(0)}$ by $1/4\pi\:\eta_{\mu\mu'}$.
On the other hand, in the case of delta
function with derivatives, we can integrate by parts
 with respect to either $\sigma$ or $\sigma'$ to transfer all the derivatives
on the prefactor in front of $H$. If any of the derivatives act on
$\tilde{G}^{(0)}$, it again follows from eq.(15) that the result either
vanishes or is proportional to an inverse power of $\epsilon$. These are the
cutoff dependent contributions and they can therefore safely be dropped.
The only term that survives is the one where all the derivatives act on
$\vessup$, and $\tilde{G}^{(0)}$ is multiplied by a delta function without 
derivatives. In this term, $\tilde{G}^{(0)}$ can again be replaced by 
$1/4\pi$, and the derivatives acting on $\vessup$ can then be shifted back on
 $H$. All this amounts to simply replacing $\tilde{G}^{(0)}$ by
$1/4\pi\:\eta_{\mu\mu'}$. As a result, we arrive at a cutoff independent
relation:
\be
b'\ids\:v(\ssp)\pdp X^{\mu\sigma}\:\frac{\delta S^{(1)}}
{\delta X^{\mu\sigma}}
+\frac{1}{16\pi}\ids\int d^{2}\sigma'\:\vessup
\frac{\delta^{2}S^{(1)}}{\delta X^{\mu\sigma}\delta X^{\mu\sigma'}}
\cong 0.
\ee
Although cutoff dependent terms have disappeared, there are still
terms belonging to the first set that have to be eliminated. This explains
the need for the $\cong$ sign; the above equation is in reality an
equivalence relation modulo terms of the first set, namely, terms which
vanish on the free field equations. The source of these unwanted terms
is the structure of $S^{(1)}$, to see this, we split it into two
pieces:
$$
 S^{(1)}= S_{f}+S_{s},
$$
where $S_{f}$ can be written in the same form as (55),
\be
S_{f}= \sum_{m,n}\ids\:N^{(m,n)}_{\mu\sigma}
\:\partial^{m}_{+}\partial^{n}_{-}X^{\mu\sigma},
\ee
 with m and n $\geq 1$, and $S_{s}$ is the rest. It is natural to expect
that $S_{f}$ will contribute terms of the first set to eq.(57).
We will now show that this is indeed the case by making use of the identity
\bq
& &\frac{\delta^{2}}{\delta X^{\mu\sigma}\delta X^{\mu\sigma'}}
\left( \sum_{m,n}\int d^{2}\tau\:N^{(m,n)}_{\lt}\:\partial^{m}_{+}
\partial^{n}_{-} X^{\lt}\right)\nonumber\\
& &\sum_{m,n}\int d^{2}\tau\:\frac{\delta^{2}N^{(m,n)}_{\lt}}
{\delta X^{\mu\sigma}\delta X^{\mu\sigma'}}\:\partial^{m}_{+}
\partial^{n}_{-} X^{\lt}+ \sum_{m,n}\partial^{m}_{\sigma_{+}}
\partial^{n}_{\sigma_{-}}(\cdots)\nonumber\\
& &+\partial^{m}_{\sigma'_{+}}\partial^{n}_{\sigma'_{-}}(\cdots).
\eq
We have not written out the explicitly
 the terms represented by dots, since they do not contribute to eq.(57).
To see this, consider the term with derivatives with respect to $\sigma_{\pm}$.
There is at least one derivative each with respect to $\ssp$ and $\sm$.
Upon integration by parts, this will kill the factor $\vessup$. A similar
argument goes through for terms
 derivatives with respect to $\sigma'_{\pm}$,
and also for the equation which is the $\sm$ counterpart of (57). The
remaining term  clearly belongs to the first set.
  It is therefore justified to drop the contribution of $S_{f}$
altogether, and replace $S^{(1)}$ by $S_{s}$ in (57). There is, however,
an ambiguity in the seperation of $S^{(1)}$ into $S_{f}$ and $S_{s}$ that
we have just outlined. This seperation depends on the form of the
integrand, as in eq.(58), and partial integration may convert an integrand
that appears to belong to the second set into one of the first set. One
way to eliminate this ambiguity is to completely fix the linear gauges
generated by integration by parts (eq.(3)), as we have done in writing
eq.(42). This is not a practical procedure in the case of higher levels,
so, we shall leave this gauge ambiguity unfixed for the time being. Later,
we shall see that, in the string language, it corresponds to the gauge
transformations generated by $L_{-1}$ and $\bar{L}_{-1}$.

Eq.(57) has exactly the same form as the linear part of renormalization
group equation derived in reference [10]. The only difference,
but an important one, is that, $S^{(1)}$ has to be replaced by $S_{s}$. This
replacement gets rid of spurious
 terms which vanish upon imposing the free field equations (45).
However, the non-linear terms in the equation derived in this paper, eq.(23),
appear to be different from the quadratic interaction term given in [10].

The next step is to translate eq.(57), with $S^{(1)}$ replaced by $S_{s}$,
into the string language, in order to compare with (56). We shall write
the analogue of eq.(1) for $S_{s}$ in the form
\be
S_{s}=\ids\:|s,\sigma>,
\ee
where the integrand is represented by a state labeled by s, which can be
built from the ``vacuum'' by applying creation operators. These operators
stand for the derivatives of X with respect to $\ssp$ and $\sm$:
\be
\partial^{m}_{+} X^{\mu\sigma}\leftrightarrow \alpha^{\dagger\mu}_{m},\;\;\;
\partial^{n}_{-} X^{\mu\sigma}\leftrightarrow \bar{\alpha}^{\dagger\mu}_{n}.
\ee
The sigma dependence of the operators $\alpha$ and $\bar{\alpha}$ has been
suppressed to simplify writing. For example, the state corresponding to
the graviton (eq.(28)) is
\be
|s,\sigma>=\tilde{h}_{\mu\nu}(X(\sigma))\:\alpha^{\dagger\mu}_{1}
\bar{\alpha}^{\dagger\nu}_{1}|0>.
\ee
We note that, since by definition, no mixed derivatives of $X$,
 such as $\pdp\dm X^{\mu\sigma}$, appear in $S_{s}$, one can write the
most general $S_{s}$ in terms of the operators defined above. The
situation here closely parallels the standard quantization of the modes
of the free string. There is, however, a difference in the way the integers
m and n in eq.(61) are assigned; in the standard string quantization, these
would stand for the Fourier modes of $X$.  Here, they represent the number of
 derivatives acting on $X$, more in parallel with the representation of the
vertex operator.

We will now rewrite the linear gauge transformations in the operator
language, by noticing that the derivatives with respect to $\ssp$ and $\sm$,
acting on a state, can be represented by
\bq
\partial_{\ssp}\rightarrow L_{-1}&=&\alpha^{\dagger\mu}_{1}\partial_{\mu}
+\sum^{\infty}_{m=1}\alpha^{\dagger\mu}_{m+1}\alpha_{m,\mu},\nonumber\\
\partial_{\sm}\rightarrow \bar{L}_{-1}&=&\bar{\alpha}^{\dagger\mu}_{1}
\partial_{\mu}+\sum^{\infty}_{n=1}\bar{\alpha}^{\dagger\mu}_{n+1}
\bar{\alpha}_{n,\mu},
\eq
and so (3) can be rewritten as
\be
|s,\sigma>\rightarrow |s,\sigma>+ L_{-1}|s_{+},\sigma>+
\bar{L}_{-1}|s_{-},\sigma>.
\ee
In the above equations,
 $\partial_{\mu}$ acts on the argument $X(\sigma)$ of a
 wavefunction such as $\tilde{h}$ in eq.(62).
 We have also introduced  annihilation operators with
 the standard commutation relations
$$
[\alpha^{\mu}_{m},\alpha^{\dagger\nu}_{n}]= \eta^{\mu\nu}\:\delta_{m,n},
\;\;\;[\bar{\alpha}^{\mu}_{m},\bar{\alpha}^{\dagger\nu}_{n}]=
\eta^{\mu\nu}\:\delta_{m,n}.
$$
Eq.(63) for $L_{-1}$ and $\bar{L}_{-1}$ is not the familiar one given in
string theory, but one can recover the standard form  by the following
scaling which preserves the commutation relations:
\be
\alpha^{\mu}_{m}=\frac{1}{\sqrt{2m}\:(m-1)!}\: a^{\mu}_{m},\;\;\;
\alpha^{\dagger\mu}_{m}=\sqrt{2m}\:(m-1)!\: a^{\dagger\mu}_{m},
\ee
and similarly for the barred operators. Here, we are guilty of an abuse of
notation;  $a$ and $a^{\dagger}$ are Hermitian conjugates, as the notation
indicates, whereas $\alpha$ and $\alpha^{\dagger}$ are not. In what
follows, we will nevertheless continue using the $\alpha$'s, since the
resulting formulas look somewhat simpler.

With these preliminaries over, we will first recast the first term in (57)
into the operator language. A straightforward calculation gives
\bq
&&\ids\:v(\ssp)\pdp X^{\mu\sigma}\frac{\delta S_{s}}{\delta X^{\mu\sigma}}
\nonumber\\
&&=\ids\Big(\sum^{\infty}_{m=1}\sum^{m}_{k=1}v^{(k)}(\ssp)
\frac{m!}{k!(m-k)!}\alpha^{\dagger\mu}_{m-k+1}\alpha_{m,\mu}
- v'(\ssp)\Big) |s,\sigma>\nonumber\\
&&=\ids\:v'(\ssp)\sum^{\infty}_{m=1}\sum^{m}_{k=1}\Big((-1)^{k-1}
\frac{m!}{k!(m-k)!}L^{k-1}_{-1}\alpha^{\dagger\mu}_{m-k+1}
\alpha_{m,\mu}\nonumber\\
&&-1\Big) |s,\sigma>
\eq
where $v^{(m)}$ stands for the m'th derivative of v with respect to its
argument. The last step follows upon integration by parts  and by replacing
the derivatives with respect to $\ssp$ by $L_{-1}$, as in eq.(63).

We now turn our attention to the second term in (57), and convert the
integrand of this term into an operator expression:
\bq
&& \frac{\delta^{2} S_{s}}{\delta X^{\mu\sigma}\delta X^{\mu\sigma'}}
=\left(\delta^{2}(\sigma-\sigma')\Box+\sum_{m}(\partial^{m}_{\ssp}
\delta^{2}(\sigma-\sigma'))\alpha^{\mu}_{m}\:\partial_{\mu}\right)|s,\sigma>
\nonumber\\ +&&\sum_{m}(-1)^{m}\partial^{m}_{\ssp}\left(\left(\delta^{2}
(\sigma-\sigma')\alpha^{\mu}_{m}\partial_{\mu}+\sum_{k}\partial^{k}_
{\ssp}(\delta^{2}(\sigma-\sigma'))\alpha^{\mu}_{m}\alpha_{k,\mu}\right)
|s,\sigma>\right)\nonumber\\
&& +\partial_{\sm}(\cdots)+\partial_{\sigma'_{-}}(\cdots)+ I_{f}.
\eq
The terms represented by dots have not been written out, since
they will drop out after
muliplication by $\vessup$ and integration over $\sigma$ and $\sigma'$.
 Also, the terms represented by $I_{f}$ belong to the
first set (eq.(55)), and as explained earlier, they do not contribute to 
the equations for the physical fields. Using this result 
and also the identity
$$
\partial^{m}_{\ssp}\partial^{k}_{\sigma'_{+}}(\vessup)_{\sigma=\sigma'}
= \frac{m!k!}{(m+k+1)!}\:v^{(m+k+1)}(\ssp),
$$
we have,
\bq
&&\ids\int d^{2}\sigma'\:\vessup\frac{\delta^{2}S_{s}}{\delta X^{\mu\sigma}
\delta X^{\mu\sigma'}}\nonumber\\
&&=\ids\Big(v'(\ssp)\:\Box+\sum^{\infty}_{m=1}v^{(m+1)}(\ssp)\:\frac{2}{m+1} 
\:\alpha^{\mu}_{m}\:\partial_{\mu}\nonumber\\
&&+\sum^{\infty}_{m=1}\sum^{\infty}_{k=1}v^{(m+k+1)}(\ssp)\:\frac{m!k!}
{(m+k+1)!}\:\alpha^{\mu}_{m}\alpha_{k,\mu}\Big) |s,\sigma>\nonumber\\
&&=\ids\:v'(\ssp)\Big(\Box+2\:\sum_{m}\frac{(-1)^{m}}{m+1}\:L^{m}_{-1}
\:\alpha^{\mu}_{m}\:\partial_{\mu}\nonumber\\
&&+\sum_{m}\sum_{k}\frac{(-1)^{m+k}m!k!}{(m+k+1)!}\:L^{m+k}_{-1}
\:\alpha^{\mu}_{m}\alpha_{k,\mu}\Big) |s,\sigma>.
\eq
In arriving at this result, again integration by parts and eq.(63) has been
used. Finally, combining eqs.(66) and (68) gives us the following
 operator version of (57): 
\be
T |s>=0,
\ee
where
$$
T= \left(\sum^{\infty}_{m=0}\frac{(-1)^{m}}{(m+1)!}\: L^{m}_{-1}L_{m}
-1\right).
$$
 For convenience, we have set $16\pi b'=1$, which differs from the
conventional slope normalization by a factor of two.
We have also introduced the conformal
operators $L_{0}$ and $L_{m}$, with $m>1$:
\bq
L_{0}&=&\Box +\sum^{\infty}_{k=1} k \alpha^{\dagger\mu}_{k}
\alpha_{k,\mu}-1,\nonumber\\
L_{m}&=&\frac{2}{m!}\:\alpha^{\mu}_{m}\:\partial_{\mu}+\sum^{m-1}_{k=1}
\frac{k!(m-k)!}{m+1}\:\alpha^{\mu}_{m-k}\alpha_{k,\mu}\nonumber\\
&+&\sum^{\infty}_{k=1}\frac{(k+m)!}{(k-1)!}\:\alpha^{\dagger\mu}_{k}
\alpha_{k+m,\mu}.
\eq
Converting the $\alpha$'s into the $a$'s through eq.(65), the L's defined
above are readily identified with the usual Virasoro operators of
string theory, with the standard commutation relations
\be
[L_{m},L_{n}]=(m-n)L_{m+n}+\frac{c}{12}(m^{3}-m)\delta_{m+n,0}.
\ee

Eq.(69) is the main result of this section. It is 
 to be supplemented by its left moving counterpart, where
$\alpha$'s are replaced by $\bar{\alpha}$'s and $L$'s by $\bar{L}$'s.
These two equations are then the linearised form of the string field
equations. It is easy to check directly that they are invariant under
the gauge transformations of eq.(64); this follows from the identity
\be
T L_{-1}=0,
\ee
where T is the operator defined in eq.(69). In their present form, these
equations ((69) and its left moving counterpart) look quite different
from the standard string equations (56); for one thing, there are only
two equations instead of an infinite set. In the next section, we will
show that, by a suitable gauge fixing, the standard string equations follow
 from (69).
\vskip 9pt
\noindent {\bf 6. Derivation Of The Standard String Equations}
\vskip 9pt

The goal of this section is to show that the  equations with $L$'s in  (56)
follow from eq.(69). Since the derivation of the equations with $\bar{L}$'s is
exactly the same, we will not consider them any further in this section.
Let us first write (69) in the form
$$
(L_{0}-1)|s>= L_{-1}|s'>.
$$
Since the right hand side of this equation is pure gauge, the left hand
side will also be pure gauge, except for states satisfying
\be
(L_{0}-1)|s>=0,
\ee
for which $L_{0}-1$ is not invertible. This is the mass shell condition
for physical states, which amounts to a partial
 choice of gauge. We note that a
rectricted set of gauge transformations, which preserve the mass shell
condition, are  still allowed. Combining (69) with (73) gives
\be
U|s>=\sum^{\infty}_{m=1}\frac{(-1)^{m}}{(m+1)!}\:L^{m-1}_{-1}L_{m}
|s>=0.
\ee
In arriving at this equation, we have used the fact that, if
$$
L_{-1}|>=0,
$$
acting on a state $|>$, then that state must vanish. Let us now grade the
states  by the eigenvalues n of the number
operator
$$
N=\sum^{\infty}_{m=1}m\:\alpha^{\dagger\mu}_{m}\alpha_{m,\mu}.
$$
The eigenvalues are then the level numbers.
The advantage of labeling the states by the level number
 follows from the fact that N commutes with
T, and so, the states with different level numbers satisfy
seperate equations of the form (69).
We will now work out the consequences of this equation, or, equivalently, of
 eqs.(73) and (74) for a few small values of n, starting with $n=0$.
The  state $|0>$  represents the tachyon, it is annihilated by all the
$\alpha$'s, and it corresponds to $p^{2}=-1$. For the next state, at
$n=1$ and $p^{2}=0$, eq.(74) gives the constraint
$$
L_{1}|1>=0.
$$
When combined with their left moving (barred) counterparts, these are
then the full set of string equations for the massless states.

Before going on to the next level, we observe that, since $L_{1}$ and
$L_{-1}$ are mutually adjoint operators, any state can be decomposed as
\be
|s>= |s'>+L_{-1}|s''>,\;\;\; L_{1}|s'>=0.
\ee
Using the gauge freedom (eq.(64)), it is therefore always possible to 
impose the condition
$$
L_{1}|s>=0,
$$
on any state. If we impose this condition on the first massive level at
$n=2$, then eq.(74), which in this case is
$$
(L_{1}-\frac{1}{3}L_{-1}L_{2})|2>=0,
$$
gives
$$
L_{2}|2>=0.
$$
We have thus recovered the full set of string equations for the level at
$n=2$.

Up to this point, the situation has been relatively simple, but at the
next level at $n=3$, new technical problems arise. Using the string
language, what we need is the
decomposition of an arbitrary state into a ``physical'' state, plus a
number of ``spurious'' states [24,25]. This decomposition,
a generalization of (75), reads
\be
|s>=|p>+|sp>.
\ee
The physical state $|p>$ satisfies the subsidiary conditions of eq.(56);
it is annihilated by all $L_{n}$'s with $n\geq 1$. On the other hand, the
spurious states $|sp>$ are formed by applying the product of various powers
of $L_{-n}$'s, again with $n\geq 1$, on a physical state. Since the $L$'s
do not commute, it is convenient to order this product according to
increasing values of n. A general spurious state can be written as
$$
|sp>=L^{n_{1}}_{-1}L^{n_{2}}_{-2}L^{n_{3}}_{-3}\cdots |>,
$$
where the state $|>$ on the right is annihilated by all $L_{n}$'s with
$n\geq 1$. In our case, we can set $n_{1}=0$, since the terms with
$n_{1}\geq 1$ can be eliminated by the gauge transformation generated by
$L_{-1}$ (eq.(64)). After this gauge fixing, the decomposition (76) can
be written as
\bq
|s,n>&=&|p,n>+\Big(c_{1} L^{n_{2}}_{-2}L^{n_{3}}_{-3}L^{n_{4}}_{-4}
\cdots + c_{2} L^{n'_{3}}_{-3}L^{n'_{4}}_{-4}L^{n_{5}}_{-5}\cdots\nonumber\\
&+&c_{3} L^{n''_{4}}_{-4} L^{n''_{5}}_{-5}\cdots+ \cdots\Big) |m>
+(\tilde{c}_{1} L^{\tilde{n}_{2}}_{-2}\cdots +\cdots) |\tilde{m}>
+\cdots
\eq
In this equation, the c's are constants, and the integers n,m, etc. are the
level numbers of the states. The right hand side is a sum over all the
possible spurious states with level number n. We are now going to apply
eq.(74) to the state $|s,n>$. It is easily verified that the physical
state $|p,n>$ satisfies this equation, and
 if we could show that all the c's must vanish,
 then we would have reached
 our goal of establishing the string equations (56). Before tackling the
general problem,
 as a simple example, let us
now consider the case $n=3$,
\be
|s,3>=|p,3>+c_{1}L_{-2}|1>+c_{2}L_{-3}|0>.
\ee
 The  state $|p,3>$ satisfies
eq.(74), and because $|1>$ and $|2>$ satisfy the physical state
 conditions and they have different level numbers, each one must satisfy
a seperate equation:
\bq
U L_{-2}|1>&=& \left(-\frac{1}{2}L_{1}+\frac{1}{6}L_{-1}L_{2}\right)
L_{-2}|1>=0,\nonumber\\
 U L_{-3}|0>&=& \left(-\frac{1}{2}L_{1}+\frac{1}{6}L_{-1}L_{2}
-\frac{1}{24} L^{2}_{-1}L_{3}\right) L_{-3}|0>=0.
\eq
Using the algebra the L's satisfy (eq.(71)), the second equation can be
 simplified to 
$$
\left(4\:L_{-2}+(\frac{c}{6}-\frac{8}{3})L^{2}_{-1}\right) |0>=0.
$$
This equation is clearly impossible, since the two states are linearly
independent and they cannot add up to zero. Hence, we must set
$c_{2}=0$ in (78). Similarly, the first equation in (79) gives
$$
(c-26) L_{-1}|1>=0.
$$
Away from the critical dimension $c=26$, we can conclude that $c_{1}=0$,
reaching our goal. However, at $c=26$, there is an ambiguity in the
definition of the physical state; it is possible to add to it a multiple
of the state $L_{-2}|1>$. We note that this state is gauge equivalent to
the zero norm state
$$
|z>= \left(L_{-2}+\frac{3}{2} L^{2}_{-1}\right)|1>,
$$
and that $|z>$ satisfies the string equations (56). The conclusion is that,
even though the physical state is not unique, nevertheless eq.(69) and 
gauge invariance under (64) still imply the standard string equations.
The possibility of adding zero norm states to a physical state is well
known from the theory of the critical string [24,25].

We would like now to apply the experience gained by working out these
special cases to the general expansion (77). Let us first sketch our
strategy. As we have noticed in working out the examples, eq.(74)
applied to (77) gives rise to several seperate equations; one for each
different state $|m>$, $|\tilde{m}>$, etc. The idea is to pick a generic
equation, and try to isolate a term from it which has a different
structure from the rest of the terms.  Such a term has to vanish
all by itself. The next step is to iterate this procedure and construct
an inductive argument. What follows is an outline of the various steps
of the argument:
 We first apply the operator U (eq.(74))
 to the right hand side of (77),
and rearrange the products over powers of $L_{-n}$'s so that n increases
from left to right, just as in (77). This is done using the commutation
relations of L's to move $L_{n}$'s for positive n to the right till they
hit the state $|m>$ and annihilate it. The result is a complicated sum
with many terms; however, one term among all others is easy to isolate;
it comes only from the application of the first term in U,
 $-\frac{1}{2}L_{1}$, to
the first term in (77):
\bq
U|s,n>&\simeq&-\frac{1}{2} c_{1}\: L_{1}L^{n_{2}}_{-2}L^{n_{3}}_{-3}
\cdots |m>\nonumber\\
&=&\left(-2\:n_{3}\: c_{1}\: L^{n_{2}+1}_{-2} L^{n_{3}-1}_{-3}\cdots +\cdots
\right) |m>.
\eq
 Since this contribution has to vanish all by itself, we conclude that
$n_{3}=0$. After setting $n_{3}=0$ in (77), we next isolate a term of
the form
\bq
U|s,n>&\simeq&\frac{1}{6} c_{1}\:L_{-1}L_{2}L^{n_{2}}_{-2}L^{n_{4}}_{-4}
\cdots |m>\nonumber\\
&=&\left( n_{4}\: c_{1}\:L_{-1}L_{-2}^{n_{2}+1} L^{n_{4}-1}_{-4}\cdots
+\cdots\right) |m>.
\eq
This term, which is again unique, is generated by the application of the
second term in U to the first term in the expansion of $|s,n>$. Since it
cannot be cancelled by any other term, it must again
 vanish all by itself, leading
to the result that $n_{4}=0$. Continuing this line of reasoning, it is easy
to show that all the n's except for $n_{2}$ must vanish. We can therefore
rewrite eq.(77) in the following form:
\be
|s,n>=\left( c_{1}\: L^{n_{2}}_{-2}+ c_{3}\: L^{n_{2}-1}_{-2} L_{-4}+\cdots
\right) |m> +\cdots .
\ee
We note that the form of the non-leading terms are severely restricted,
since  their grading with respect to the number operator must match the
grading of the leading term, which is $2\:n_{2}$. Again applying U to
$|s,n>$ given above, we identify two terms which must vanish individually:
\bq
U|s,n>&\simeq& \Big(-\frac{1}{2}L_{1}+\frac{1}{6}L_{-1}L_{2}\Big)
|s,n>\nonumber\\
&\simeq& \bigg(\Big(-\frac{1}{2}(3\:n_{2}+\frac{4}{3} n^{2}_{2}
-\frac{1}{6}c\: n_{2})c_{1} + c_{3}\Big) L_{-1} L^{n_{2}-1}_{-2}\nonumber\\
&+& \Big(\frac{3}{4}n_{2}(n_{2}-1) c_{1}- \frac{5}{2}\:c_{3}\Big)
L^{n_{2}-2}_{-2} L_{-3}\bigg) |m>.
\eq
 The resulting equations
$$
3\:n_{2}(n_{2}-1) c_{1}- 10\:c_{3}=0,
$$
and
$$
(3\:n_{2}+\frac{4}{3}n^{2}_{2}-\frac{1}{6} c\: n_{2}) c_{1}
- 2\:c_{3}=0,
$$
have the only non-trivial solution $n_{2}=1$, $c_{3}=0$,
in the critical dimension $c=26$. This solution, which can be absorbed into
into the physical state by a redefinition, was discussed following eq.(78).
We can therefore conclude that
$$
c_{1}=0,\;\;\; c_{3}=0.
$$
Incorporating this result into the expansion (77) gives
\be
|s,n>=\left( c_{4}\: L_{-3}^{n_{3}} L^{n_{4}}_{-4}\cdots + c_{5}\:
L^{n_{3}-1}_{-3} L^{n'_{4}}\cdots +\cdots \right)|m>.
\ee
The absence of factors which contain $L_{-2}$ simplies greatly the next
step in the argument. We again isolate a unique term from eq.(74):
\bq
U|s,n>&\simeq& -\frac{1}{2}  L_{1}\left( c_{4}\:L^{n_{3}}_{-3} L^{n_{4}}_{-4}
\cdots+\cdots\right) |m>\nonumber\\
&\simeq& \left(-2\: n_{3}\: c_{4}\: L_{-2}L_{-3}^{n_{3}-1}\cdots +\cdots
\right) |m>.
\eq
For this term to vanish, $n_{3}$ must be equal to zero. This line of resoning
can be continued inductively to show that all the spurious states on the
right hand side of eq.(77) are absent, leaving behind only the physical
state. Since the physical states by definition satisfy the string field
equations of (56), we have therefore succeded in deducing these equations
from eq.(69).
\vskip 9pt
\noindent {\bf 7. Gravity To Higher Orders}
\vskip 9pt

So far, we have only studied the linearized form of the string field
equations. In this section, we will consider higher order interaction
terms of the massless fields. To keep the discussion simple, we restrict
 ourselves to a symmetric metric, with $\tilde{h}_{\mu\nu}=h_{\mu\nu}=
h_{\nu\mu}$ in eq. (28), and drop the antisymmetric tensor $B$,
 although there is no real difficulty in treating
the general case. 
We have already shown in section 3 that the linear terms in the equations
of motion are those of gravity coupled to a dilaton,
 and in view of the postulated invariance 
under coordinate transformations (30), one would expect that the full
non-linear set of equations
 will also turn out to be Einstein's equations for the
dilaton-graviton system. We think, nevertheless, that it is worthwhile 
to verify that the field equations come out correctly for several reasons.
For one thing, it is important to
 demonstrate that the scheme of regularization we
are using respects the covariance under coordinate transformations. 
Also, it is nice to have
 a check on the method of using an initially
undetermined connection. In the linear approximation, it turned out that 
the connection could be completely eliminated from the final field
equations, leaving behind only the contribution of the dilaton field.
We would like to verify that the same thing continues to hold true when
the higher order terms are taken into account. Finally, the computation
carried out in this section is a necessary preliminary to an explicit
construction of a full set cutoff independent non-linear equations for all
the levels of the string. Although this latter problem is not considered
in this paper, we hope to return to it in a future publication.

The starting point is eq.(23), with the action given by eq.(28) and a
symmetric $\tilde{h}_{\mu\nu}=h_{\mu\nu}$.
 We can simplify this equation  simplify using
eq.(32) and setting
$f^{\mu\sigma}_{\sigma'}=0$. With these simplifications, eq.(23) becomes
\be
W^{(1)}+W^{(2)}+E_{M}=0,
\ee
where,
\bq
W^{(1)}&=&\frac{1}{4}\ids\int d^{2}\sigma'\vessup\:\epsilon\:
\tilde{G}^{(0)}_{\mu'\sigma',\mu\sigma}(\epsilon) H_{\mu\sigma,
\mu'\sigma'},\nonumber\\
W^{(2)}&=&-\frac{1}{8}\ids\int d^{2}\sigma'\vessup\nonumber\\
&\times& \int^{\epsilon}_
{0}dt'(\epsilon - t')\tilde{G}^{(0)}_{\mu'\sigma',\mu\sigma}(\epsilon
-t')\left(H \tilde{G}(t') H\right)_{\mu\sigma,\mu'\sigma'}.
\eq
$W^{(1)}$, the linear term, was already computed in section 3, so we turn
our attention to the second term, $W^{(2)}$, which contains all the
non-linear contribution . In order to keep the exposition
simple, we will present here only the details of the computation of the
quadratic terms in the fields in $W^{(2)}$, although it not too difficult
to treat higher order terms by the same method we are using. We will also 
not carry out the calculation of the determinental term $E_{M}$ to higher
orders. Just as in the linear case (eq.(34)), $E_{M}$ turns out to be
purely cutoff dependent also in the higher orders, and it is cancelled
by the cutoff dependent parts of $W^{(1)}$ and $W^{(2)}$. This is the same
as computing the higher order corrections to eq.(36), and since we are
only interested in computing the higher order contributions to eq.(35),
we will not consider $E_{M}$ any further.

 The part of
$W^{(2)}$ quadratic in the fields is given by
\bq
W^{(2)}&\simeq& -\frac{1}{8}\ids\int d^{2}\sigma'\vessup\nonumber\\
&\times& \int^{\epsilon}_
{0}dt' (\epsilon - t') \tilde{G}^{(0)}_{\mu'\sigma',\mu\sigma}(\epsilon
- t') \left(H \tilde{G}^{(0)}(t') H\right)_{\mu\sigma,\mu'\sigma'}.
\eq
From its definition (first line of eq.(27)), $H$ can be expressed in terms
of the metric and the connection:
\bq
H_{\mu\sigma,\mu'\sigma'}&=&\partial_{\ssp}\partial_{\sm}\left(
\delta^{2}(\sigma - \sigma') A_{\mu\mu'}\right)\nonumber\\ 
&+&\partial_{\ssp}\left(\delta^{2}(\sigma - \sigma') \dm X^{\lambda\sigma}
A_{\mu\mu',\lambda}\right)+ \partial_{\sm}\left(\delta^{2}(\sigma - \sigma')
\pdp X^{\lambda\sigma} A_{\mu\mu',\lambda}\right)\nonumber\\
&+& \delta^{2}(\sigma - \sigma')\left(\pdp X^{\lambda\sigma}
\dm X^{\lambda'\sigma}\:B_{\mu\mu',\lambda\lambda'}+ \pdp\dm X^{\lambda
\sigma}\:B_{\mu\mu',\lambda}\right),
\eq
where the A's and the B's are given by
\bq
A_{\mu\nu}&=&-2\:h_{\mu\nu},\nonumber\\
A_{\mu\nu,\lambda}&=&\partial_{\lambda}h_{\mu\nu}-\partial_{\nu}h_{\mu\lambda}
+\partial_{\mu}h_{\nu\lambda},\nonumber\\
B_{\mu\nu,\lambda\lambda'}&=&\partial_{\mu}\partial_{\nu}h_{\lambda\lambda'}
-\partial_{\mu}\partial_{\lambda}h_{\nu\lambda'}-\partial_{\mu}
\partial_{\lambda'}h_{\nu\lambda}\nonumber\\
&+&\Gamma^{\eta}_{\mu\nu}(\partial_{\lambda}h_{\eta\lambda'}+
\partial_{\lambda'}h_{\eta\lambda}-\partial_{\eta}h_{\lambda\lambda'}),
\nonumber\\
B_{\mu\nu,\lambda}&=&-2\:\partial_{\mu}h_{\nu\lambda}+2\:\Gamma^{\eta}_
{\mu\nu}\:g_{\eta\lambda}.
\eq
Substituting this in eq.(85) expresses the quadratic contributions in terms of
the metric and the connection. This is not the end of the story, however,
since we still have to extract the finite and cutoff dependent terms from
this expression in the limit of $\epsilon \rightarrow 0$. We observe that  the
cutoff dependendence comes from a factor of the form
\be
D^{(m,n)}_{\sigma,\sigma'}(\epsilon)= \int^{\epsilon}_{0}dt'(\epsilon - t')
\tilde{G}^{(0)}_{\sigma',\sigma}(\epsilon - t') \partial^{m}_{\ssp}
\partial^{n}_{\sm}\left(\tilde{G}^{(0)}_{\sigma,\sigma'}(t')\right).
\ee
The derivatives acting on the second $\tilde{G}$ come from the derivatives of
delta functions in the expression for $H$ (eq.(86)), and so
$0\leq m \leq 2$, $0 \leq n \leq 2$. From eq.(15), it can easily be shown
that, in the limit of $\epsilon \rightarrow 0$, the D's either vanish or
tend to various derivatives of delta functions. Below is a list of the
\underline{only non-vanishing}
 limits of the D's for both m and n less than or equal
to two:
\bq
D^{(1,1)}_{\sigma,\sigma'}(\epsilon)&\rightarrow& -\frac{1}{8\pi}
\delta^{2}(\sigma - \sigma'),\nonumber\\
D^{(2,1)}_{\sigma,\sigma'}(\epsilon)&\rightarrow&-\frac{1}{6\pi}
\partial_{\ssp}\delta^{2}(\sigma - \sigma'),\nonumber\\
D^{(1,2)}_{\sigma,\sigma'}(\epsilon)&\rightarrow&-\frac{1}{6\pi}
\partial_{\sm}\delta^{2}(\sigma - \sigma'),\nonumber\\
D^{(2,2)}_{\sigma,\sigma'}(\epsilon)&\rightarrow&-\frac{5}{24\pi}
\partial_{\ssp}\partial_{\sm}\delta^{2}(\sigma - \sigma')+
\frac{1}{4\pi\epsilon}\delta^{2}(\sigma - \sigma').
\eq
Putting everything together, in the limit of $\epsilon \rightarrow 0$,
$W^{(2)}$ can be written as
\be
W^{(2)}=\ids v'(\ssp) \pdp X^{\lambda\sigma}\dm X^{\lambda'\sigma}
Z^{(2)}_{\lambda\lambda'}-\frac{1}{8\pi\epsilon}\ids v'(\ssp)( h_{\mu\nu}
h_{\mu\nu}),
\ee
where,
\bq
Z^{(2)}&=&-\frac{1}{8\pi}\bigg(\frac{1}{2}\partial_{\mu}h_{\nu\lambda}
\partial_{\mu}h_{\nu\lambda'}-\frac{1}{4}\partial_{\lambda}h_{\mu\nu}
\partial_{\lambda'}h_{\mu\nu}-\frac{1}{2}\partial_{\mu}h_{\nu\lambda}
\partial_{\nu}h_{\mu\lambda'}\nonumber\\
&+&\frac{1}{2}h_{\mu\nu}(\partial_{\mu}
\partial_{\nu}h_{\lambda\lambda'}-\partial_{\mu}\partial_{\lambda}
h_{\nu\lambda'}-\partial_{\mu}\partial_{\lambda'}h_{\nu\lambda})
+\frac{1}{2}\Gamma^{(1)\eta}_{\mu\mu}(\partial_{\eta}h_{\lambda\lambda'}
-\partial_{\lambda}h_{\eta\lambda'}-\partial_{\lambda'}h_{\eta\lambda})
\nonumber\\
&+&\partial_{\lambda'}\left(h_{\mu\nu}(\partial_{\mu}h_{\nu\lambda}
-\Gamma^{(1)\lambda}_{\mu\nu})+\Gamma^{(1)\eta}_{\mu\mu}h_{\eta\lambda}
+\Gamma^{(2)\lambda}_{\mu\mu}\right)\bigg).
\eq
The first order contribution to $Z$ was calculated in section 3:
\be
Z^{(1)}_{\lambda\lambda'}= \frac{1}{16\pi}\left(\Box h_{\lambda\lambda'}
-\partial_{\lambda}\partial_{\mu}h_{\mu\lambda'}+\partial_{\lambda'}
\partial_{\mu}h_{\mu\lambda}
-2\;\partial_{\lambda'}
\Gamma^{(1)\lambda}_{\mu\mu}\right),
\ee
where $\Gamma^{(1)}$ and $\Gamma^{(2)}$ stand for the
linear and quadratic parts of the connection. The generalization of eq.(35)
to include quadratic terms is then given by
\be
Z^{(1)}_{\lambda\lambda'}+Z^{(2)}_{\lambda\lambda'}=0.
\ee
From this equation, using the symmetry of $h_{\mu\nu}$ in $\mu$ and $\nu$,
it is easy to show that
$$
\partial_{\lambda'}\left((g_{\mu\nu}g_{\eta\lambda}\Gamma^{\eta}_{\mu\nu})_
{(2)} -\partial_{\mu}h_{\mu\lambda}+h_{\mu\nu}\partial_{\mu}
 h_{\nu\lambda}\right)
-(\lambda\leftrightarrow \lambda')=0,
$$
where the subscript (2) means that only up to second order contributions
are included. The solution to this equation can be written as
\be
(g^{\mu\nu}g_{\eta\lambda}\Gamma^{\eta}_{\mu\nu})_{(2)}= \partial_{\mu}
h_{\mu\lambda} -h_{\mu\nu}\partial_{\mu}h_{\nu\lambda}-\frac{1}{2}
\partial_{\lambda}(h_{\mu\mu}- h_{\mu\nu}h_{\mu\nu})+
\partial_{\lambda}\phi,
\ee
where $\phi$ is identified with the dilaton field. There is always some
ambiguity in the definition of the dilaton field; for example, we could
have defined a different dilaton field by
$$
\bar{\phi}= \phi -\frac{1}{2}h_{\mu\mu}+\frac{1}{2}h_{\mu\nu}h_{\mu\nu},
$$
and thereby simplified eq.(93). This ambiguity arises from the well
known possibility of mixing the dilaton with the determinant of the metric.
In the definition we have chosen, the dilaton transforms as a scalar
under coordinate transformations.

Finally, the terms involving the connection in eq.(92) can be eliminated
using eq.(93). As promised earlier, to second order in h,
 the resulting field equations coincide
with the equations
\be
2\:R_{\mu\nu}+D_{\mu}D_{\nu}\phi=0,
\ee
of the gravity-dilaton system.

\vskip 9pt
\noindent {\bf 8. Conclusions}
\vskip 9pt

 In this paper, we developed further and extended the method for deriving
string field equations proposed in an earlier paper [1].
As a check on the method, we derived the linearized equations for all string
states and the full non-linear equations for the dilaton-graviton system and
compared them with the well known results. There seems to be no obstacle to
obtaining a full set of interacting equations for all levels. These
equations would then enjoy the desirable properties of background independence
and covariance under general non-local coordinate transformations.

\vskip 9pt
\noindent {\bf Acknowledgements}
\vskip 9pt

I would like to thank Luis Bernardo for discussions.
\newpage
{\bf References}
\begin{enumerate}
\item K.Bardakci and L.M. Bernardo,\emph{String Field Equations
from Generalized Sigma Model}, hep-th/9701171.
\item D.Friedan, Phys.Rev.Lett.{\bf 45}(1080) 1057.
\item L.Alvarez-Gaum\'{e}, D.Z.Freedman and S.Mukhi, Ann. of Phys.{\bf 134}
(1981) 85.
\item C.Lovelace, Phys. Lett. {\bf 135B}(1984) 75.
\item E.S.Fradkin and A.A.Tseytlin, Nucl. Phys. {\bf B261} (1985) 1.
\item C.G.Callan, D.Friedan, E.Martinec and M.J.Perry, Nucl.Phys.
{\bf B262} (1985) 593.
\item T.Banks and E.Martinek, Nucl.Phys.{\bf B294} (1987) 733.
\item K.G.Wilson and J.G.Kogut, Phys. Rep. {\bf 12} (1974) 75.
\item J.Polchinski, Nucl. Phys. {\bf B231} (1984) 413.
\item J.Hughes, J.Liu and J.Polchinski, Nucl. Phys. {\bf B316} (1989) 15.
\item A.N.Redlich, Phys. Lett. {\bf 213B} (1988) 285.
\item R.Brustein and K.Roland, Nucl.Phys. {\bf B372} (1992) 201.
\item A.A.Tseytlin, Int. J. Mod. Phys. {\bf A16} (1989) 4249.
\item U.Ellwanger and J.Fuchs, Nucl. Phys. {\bf B312} (1989) 95.
\item E.Braaten, T.L.Curtright and C.K.Zachos, Nucl. Phys {\bf B260}
(1985) 630.
\item C.B.Thorn, Phys. Rep. {\bf 175} (1989) 1.
\item M.Bochicchio, Phys. Lett. {\bf 193B} (1987) 31.
\item E.Witten, Phys. Rev. {\bf D46} (1992) 5467.
\item B.Zwiebach, Nucl. Phys. {\bf B480} (1996) 541, A.Sen and
B.Zwiebach, Nucl. Phys. {\bf B423} (1994) 580. 
\item For a review, see C.Vafa, \emph{Lectures on Strings and Dualities},
hep-th/9702201
\item I.A.Batalin and G.A.Vilkovisky, Phys. Rev. {\bf D28} (1983) 2567.
\item R.Brustein and S.de Alwis, Nucl. Phys. {\bf B352} (1991) 451.
\item M.Henneaux, in \emph{Quantum Mechanics of Fundamental Systems 3},
edited by C.Teitelboim and J.Zanelli. Plenum Press, New York, 1992.
\item M.B.Green, J.H.Schwarz and E.Witten, \emph{Superstring Theory},
Cambridge University Press, 1987. 
\item P.Goddard, C.Rebbi and C.Thorn, Nuovo Cim. {\bf 12A} (1972) 425.
\end{enumerate}
\end{document}